           \documentclass[11pt,twocolumn,a4paper]{article}

\usepackage{amssymb}
\usepackage{siunitx}
\usepackage{graphicx}
\usepackage{caption}
\usepackage[T1]{fontenc}
\usepackage[utf8]{inputenc}
\usepackage{authblk}
\usepackage{subcaption}

\usepackage[breaklinks = true, colorlinks = true, linkcolor = blue,citecolor = cyan]{hyperref}

\begin{document}




\title{Simulations of 3D-Si sensors for the innermost layer of the ATLAS pixel upgrade}


\author[1]{M.~Baselga}
\author[1]{G.~Pellegrini\thanks{giulio.pellegrini@imb-cnm.csic.es}}
 \author[1]{D.~Quirion}

\affil[1]{Centro Nacional de Microelectr\'{o}nica, CNM-IMB (CSIC), Barcelona 08193, Spain}
\maketitle

\begin{abstract}

The LHC is expected to reach luminosities up to \SI{3000}{\per\femto\barn} and the innermost layer of the ATLAS upgrade plans to cope with higher occupancy and to decrease the pixel size. 3D-Si sensors are a good candidate for the innermost layer of the ATLAS pixel upgrade since they exhibit good performance under high fluences and the new designs will have smaller pixel size to fulfill the electronics expectations. This paper reports TCAD simulations of the 3D-Si sensors designed at IMB-CNM with non passing-through columns that are being fabricated for the next innermost layer of the ATLAS pixel upgrade, shows the charge collection response before and after irradiation, and the response of 3D-Si sensors located at large $\eta$ angles.

\end{abstract}





\section{Introduction}
\label{Introduction}

The LHC is expected to reach luminosities up to \SI{3000}{\per\femto\barn} and the innermost layer of the ATLAS pixel silicon detectors will have to cope with fluences of $2 \times 10^{16}$ \SI{1}{\MeV} neutron equivalent per square centimeter (\si{n_{eq} \cm^{-2}})\cite{Gianotti2005}. The new tracking silicon detectors will have to deal with an increased occupancy for each detector, be more radiation hard, and the future 3D-Si sensors will have to fulfill the specifications of the new read out electronics developed by the RD53 collaboration\cite{electronics2013}.

3D-Si sensors \cite{Parker1997} typically have smaller distances between the electrodes and are fully depleted at lower voltages than planar ones\cite{Pennicard2007}, they have less probability of trapping once irradiated, since they typically have shorter drift distances, and they show some multiplication after irradiation\cite{Kohler2011}. 
They were first inserted inside a high energy physics experiment for the IBL (Insertable b-Layer) during the long shutdown 1 in 2013\cite{ATLASIBLCollaboration2012}. 3D-Si sensors and the planar ones show a good performance inside the IBL\cite{ATLAS2015} and they may be used for the new ATLAS phase II upgrade. The designs of the 3D-Si sensors proposed for the ATLAS pixel upgrade have a pixel size of \SI{50}{\micro\m}$\times$\SI{50}{\micro\m} to decrease the occupancy and to decrease the drift distances\cite{electronics2013}. Since the new read out electronics developed by the RD53 collaboration will not be ready until 2017, the first 3D-Si sensors with smaller pixel sizes will have to fit the actual FE-I4 chip in order to be tested under similar conditions than the final ones. The IMB-CNM design aims to increase the aspect ratio, the ratio between the depth of the column and the diameter of the column, by means of the use of the cryogenic DRIE process\cite{wu2010} to fulfill the new specifications. Pixel cells with the size of \SI{25}{\micro\m}$\times$\SI{100}{\micro\m} are also under consideration for the ATLAS phase II upgrade. 

This work reports the simulations for IMB-CNM 3D-Si sensors designed for the new ATLAS upgrade and their performance under high fluences and at large $\eta$ angles.
The detectors are fabricated in a similar method as reported in ref. \cite{Pellegrini2013b} and they will use the mask detailed in ref. \cite{Baselga2015b}.


\section{Simulations for the new ATLAS upgrade}
\label{Simulations}

 The simulations are carried out using the Synopsys Sentaurus TCAD simulation toolkit\cite{synopsis_sentaurus} for the two pixel sizes (\SI{50}{\micro\m}$\times$\SI{50}{\micro\m} and \SI{100}{\micro\m}$\times$\SI{25}{\micro\m}). The simulated detectors are \SI{200}{\micro\m} thick with a p-type substrate and the radiation trap model used is the one reported in ref. \cite{Pennicard2010a}. This model is based on the Perugia model and was not conceived for large fluences, thus the results might be an approximation. The silicon dioxide surface charge for a non-irradiated detector is $Q_{ox} = $ \SI{1e11}{\cm^{-2}} and it is expected to saturate at $Q_{ox} = $\SI{3e11}{\cm^{-2}} after irradiation\cite{Lutz1999}. The simulations were carried out at a temperature of \SI{258}{\kelvin}.

\begin{figure} [htb!]
\centering \includegraphics[width=\linewidth]{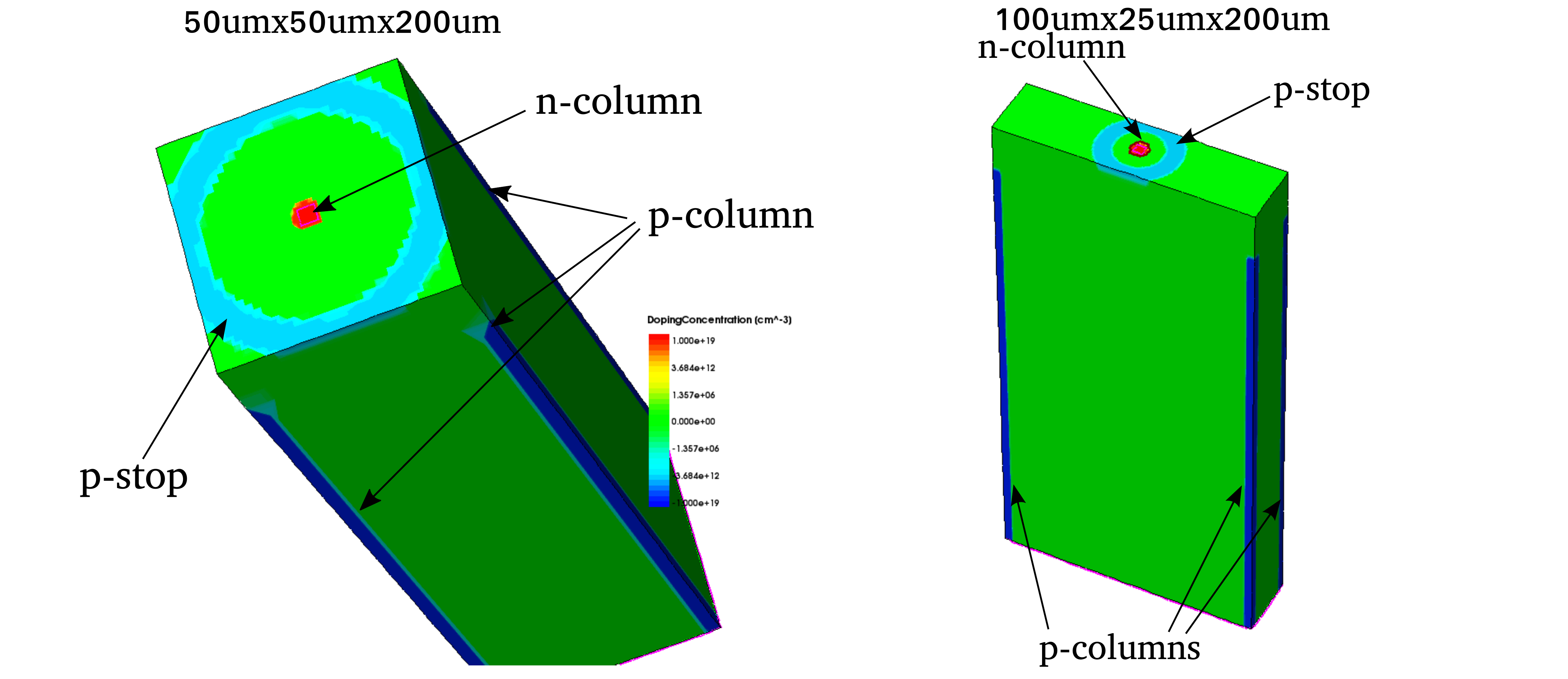}
   \caption{Simulated pixel cell for \SI{50}{\micro\m}$\times$\SI{50}{\micro\m}$\times$\SI{200}{\micro\m} (left) and \SI{100}{\micro\m}$\times$\SI{25}{\micro\m}$\times$\SI{200}{\micro\m} (right) geometry.}
   \label{dos_cols}
\end{figure}

The simulated structure is a 3D-Si with IMB-CNM double sided design similar as the one shown in ref. \cite{Pellegrini2008}, with non passing-through columns, a depth of \SI{170}{\micro\m} and diameter of \SI{5}{\micro\m}. Fig. \ref{dos_cols} shows the two simulated pixel cells, which have a p-stop radius half the pitch of each geometry, \SI{25}{\micro\m} of radius for the \SI{50}{\micro\m}$\times$\SI{50}{\micro\m} geometry and \SI{12.5}{\micro\m} of radius for the \SI{100}{\micro\m}$\times$\SI{25}{\micro\m} geometry.

\begin{figure}[h!]
\centering \includegraphics[width=\linewidth]{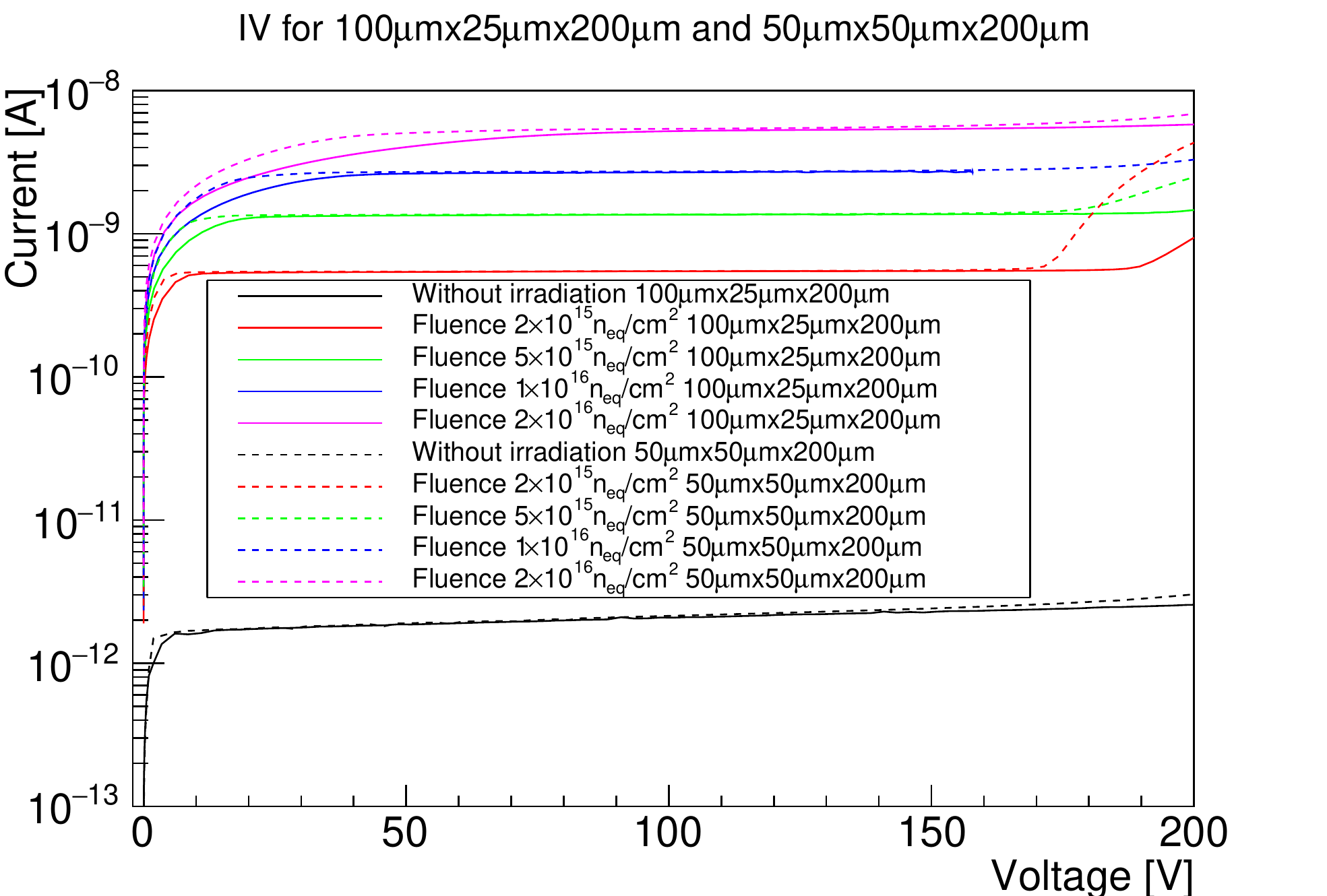}
   \caption{Current-voltage curves simulated for \SI{50}{\micro\m}$\times$\SI{50}{\micro\m}$\times$\SI{200}{\micro\m} and  \SI{100}{\micro\m}$\times$\SI{25}{\micro\m}$\times$\SI{200}{\micro\m} single pixel cells at different fluences.}
   \label{sim_iv_col5050}
       \end{figure}

Fig. \ref{sim_iv_col5050} shows the current-voltage curves for the \SI{50}{\micro\m}$\times$\SI{50}{\micro\m}$\times$\SI{200}{\micro\m} and \SI{100}{\micro\m}$\times$\SI{25}{\micro\m}$\times$\SI{200}{\micro\m} pixel size at different fluences. Both structures show similar current. In both cases it can be seen that breakdown does not not occur below \SI{200}{\V}, after \SI{200}{\V} the simulation do not converge anymore probably due to the break down. 

 The damage constant $\alpha$ is lower than the expected value due to the fact that the simulations are not well adjusted for high fluences (as pointed out in ref. \cite{Pennicard2010a}) such as \SI{2e16}{n_{eq}/\cm^2}.

\begin{figure} [htb!]
\centering \includegraphics[width=\linewidth]{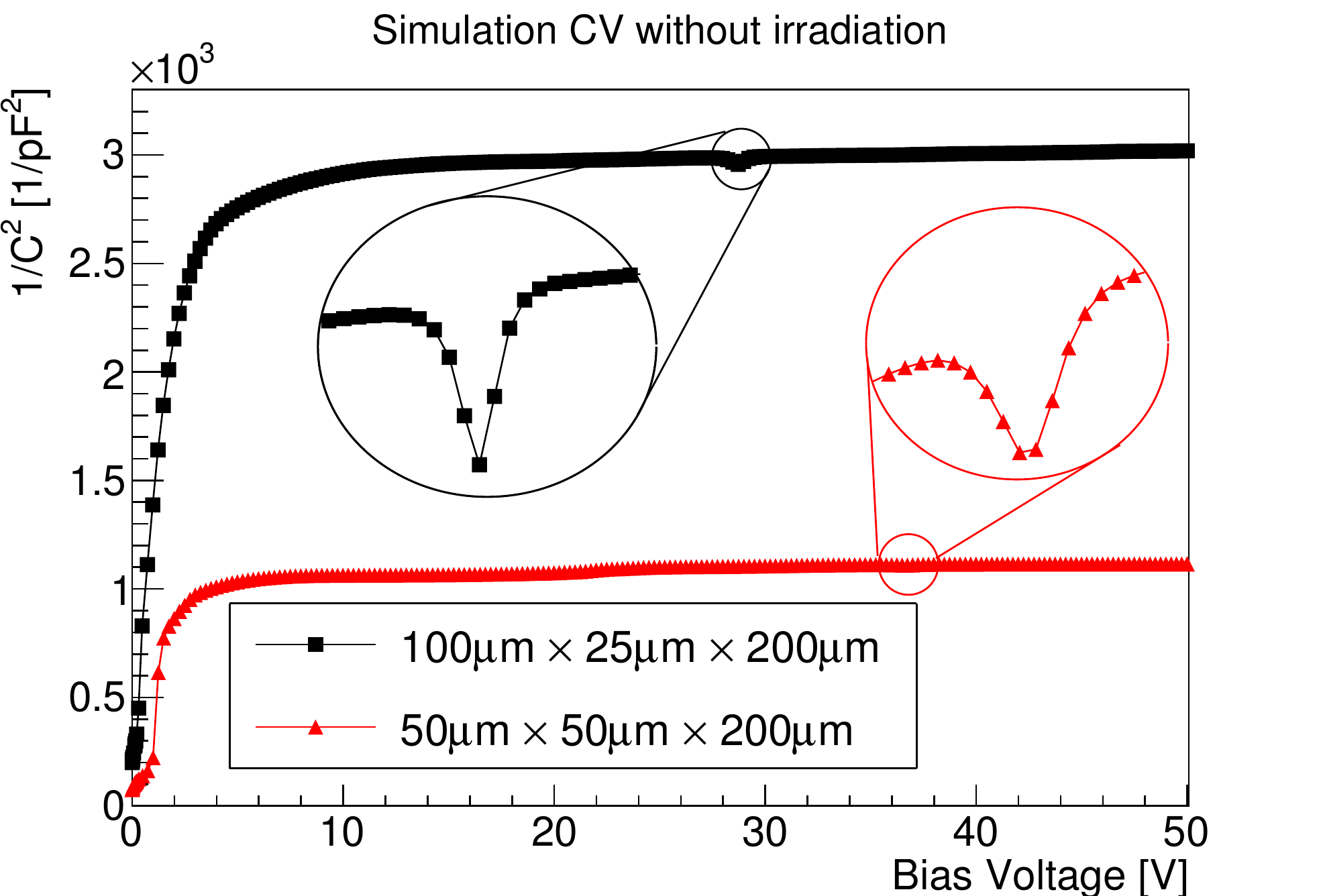}
   \caption{$1/C^2$-voltage curves for \SI{50}{\micro\m}$\times$\SI{50}{\micro\m}$\times$\SI{200}{\micro\m} and \SI{100}{\micro\m}$\times$\SI{25}{\micro\m}$\times$\SI{200}{\micro\m} geometries.}
   \label{comp_cols_edit}
\end{figure}

 \sloppy
 Fig. \ref{comp_cols_edit} shows the $1/C^2$-voltage curves of the simulated structures, and the capacitance of 
 \SI{50}{\micro\m}$\times$\SI{50}{\micro\m}$\times$\SI{200}{\micro\m} pixel size is larger than the one for 
 \SI{100}{\micro\m}$\times$\SI{25}{\micro\m}$\times$\SI{200}{\micro\m}. Both structures show a bump, at \SI{28.7}{\V} and at \SI{36.5}{\V}, respectively, where the structure is supposed to reach the full depletion\cite{Stewart2013, BalbuenaValenzuela2012}. 
  Fig. \ref{comp_cols_edit} shows a lateral depletion at $\approx$\SI{3}{\V} for \SI{50}{\micro\m}$\times$\SI{50}{\micro\m}$\times$\SI{200}{\micro\m}, which is slightly lower than the values of \SI{100}{\micro\m}$\times$\SI{25}{\micro\m}$\times$\SI{200}{\micro\m} which is $\approx$\SI{4}{\V}, as expected by the geometry. 
 
\subsection{MIP through \SI{50}{\micro\m}$\times$\SI{50}{\micro\m} pixel cell}
\label{mip5050}

\begin{figure}[h!]
\centering \includegraphics[width=\linewidth]{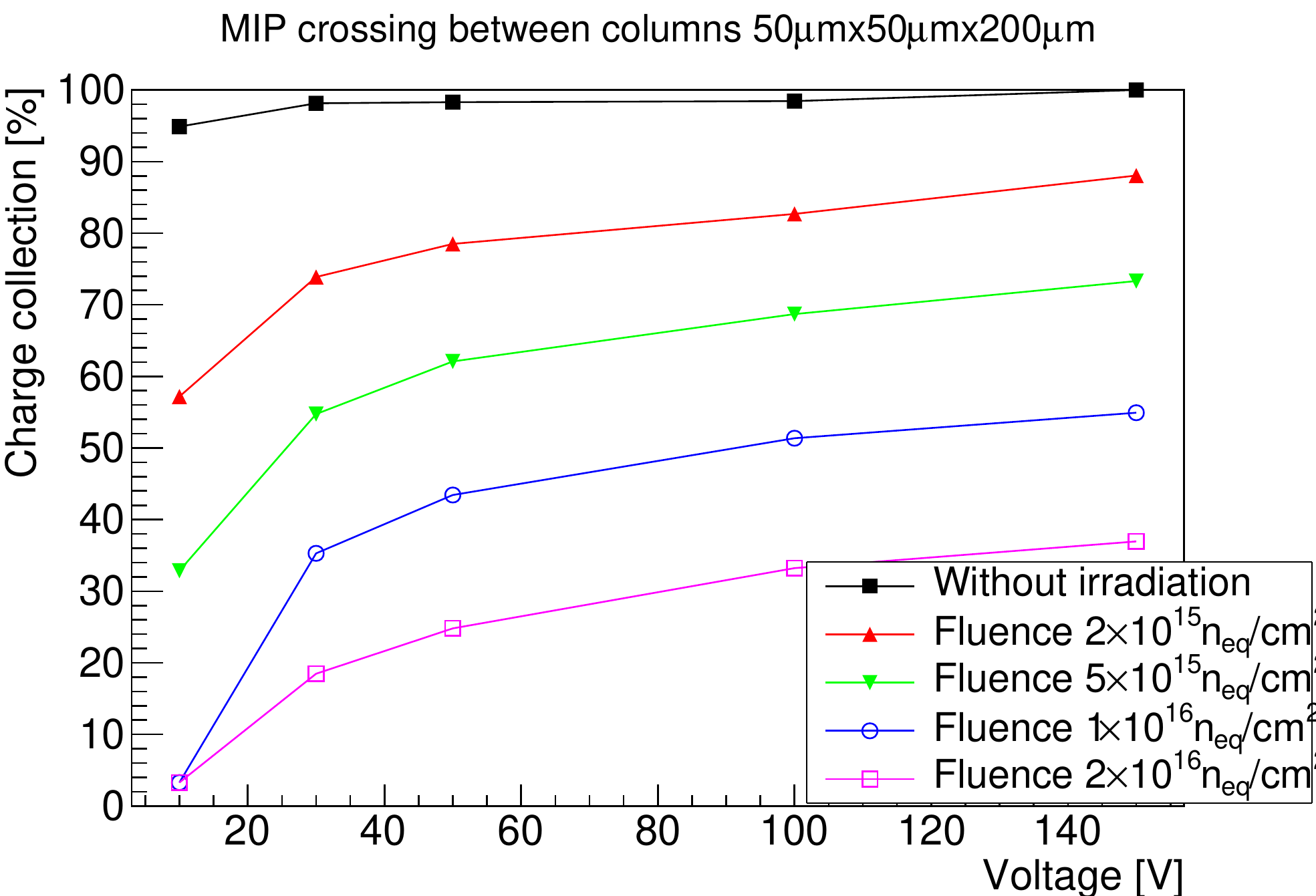}
   \caption{Simulation of the charge collection for a MIP crossing the detector between the p-column and the n-column for \SI{50}{\micro\m}$\times$\SI{50}{\micro\m}$\times$\SI{200}{\micro\m}.}
   \label{col_50_50_cc}

\end{figure}

A Minimum Ionizing Particle (MIP) was simulated with the HeavyIon function of the Synopsys Sentaurus TCAD toolkit, with a LET\_f =\SI{1.282e-5}{\pico\coulomb \per\micro\m} (Linear Energy Transfer function). 
 Fig. \ref{col_50_50_cc} shows the simulation of the collected charge for a MIP impinging perpendicular to the surface of the 3D-Si sensor (parallel to the columns) between the p-column and n-column (at a distance of $\sqrt{12.5^2+12.5^2}=$\SI{17.7}{\micro\m}) with an integration time of \SI{25}{\nano\s}. This simulation was run with a single pixel cell neglecting the charge sharing between neighboring columns. The Charge Collection Efficiency (CCE) decreases with irradiation and reaches values up to 30\% once the device is irradiated with a fluence \SI{2e16}{n_{eq} \cm^{-2}} at \SI{150}{\V}.

\begin{figure}[h!]
       \centering \includegraphics[width=\linewidth]{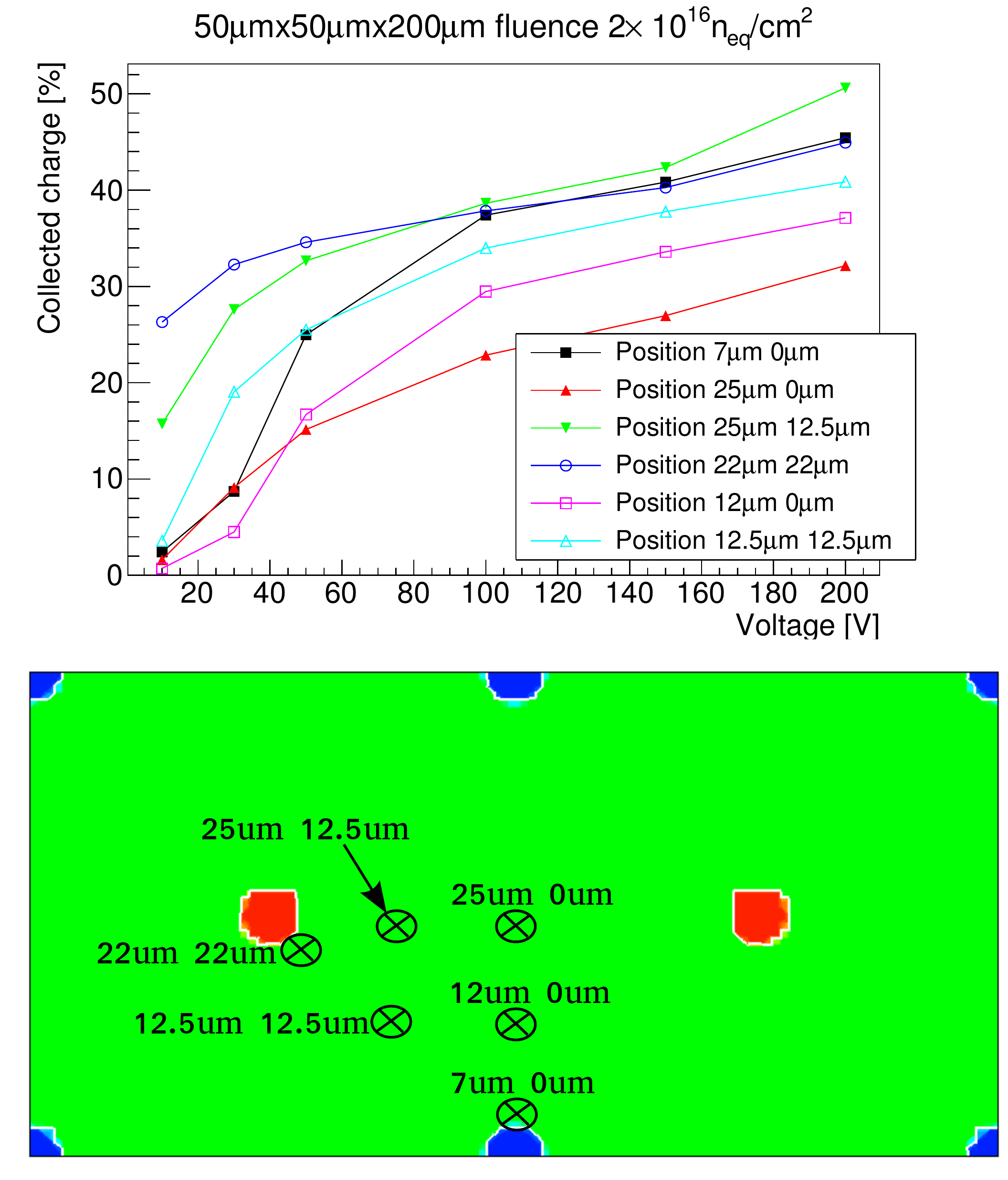}
       \caption{Collected charge for MIP in different positions for irradiated \SI{50}{\micro\m}$\times$\SI{50}{\micro\m}$\times$\SI{200}{\micro\m} sensor at \SI{2e16}{n_{eq} \cm^{-2}}. The lower part of the figure shows the different positions of the MIP. }
       \label{c5050_2e16}
\end{figure}

Fig. \ref{c5050_2e16} show the simulation of the collected charge of a MIP crossing the pixel cell at different positions (the lower part of fig. \ref{c5050_2e16} shows the simulated MIP positions) for an irradiated detector with a fluence of \SI{2e16}{n_{eq} \cm^{-2}}, the non irradiated device shows a CCE close up to 100\% for all the positions since no charge is loss in the ideal detector without traps. The integrated charge of fig. \ref{c5050_2e16} is normalized to the maximum charge of a non irradiated device (the one of the non irradiated simulation). The maximum CCE is up to 50\% when the detector is biased at \SI{200}{\V} and irradiated with a fluence of \SI{2e16}{n_{eq} \cm^{-2}}, and the minimum is at 30\%. The simulation took into account two consecutive pixel cells in order to integrate the possible charge observed by the neighboring columns, and the integration time is always \SI{25}{\nano\s} in order to agree with the HL-LHC bunch crossing. The difference in CCE after irradiation from 30\% to 50\% is due to fact that electron-hole pairs created farther from the electrode with lower electric field will percieve more trapping than electron-hole pairs created at higher elecric field regions or nearer to the electrode. Ref. \cite{Mendicino2015} reported similar results for the FBK geometry, which is a passing-through column and the simulation only took into account one quarter part of the pixel cell. The simulations reported in ref. \cite{Mendicino2015} for the \SI{100}{\micro\m}$\times$\SI{25}{\micro\m} irradiated with fluence of \SI{2e16}{n_{eq} \cm^{-2}} shows a CCE between 40\% and 60\% and the \SI{50}{\micro\m}$\times$\SI{50}{\micro\m} shows a lower CCE between 20\% and 40\% depending on the position of the MIP. The values are comparables with the ones reported in fig. \ref{c5050_2e16}.


\subsection{MIP through \SI{100}{\micro\m}$\times$\SI{25}{\micro\m} pixel cell}

\begin{figure} [htbp!]
\centering \includegraphics[width=\linewidth]{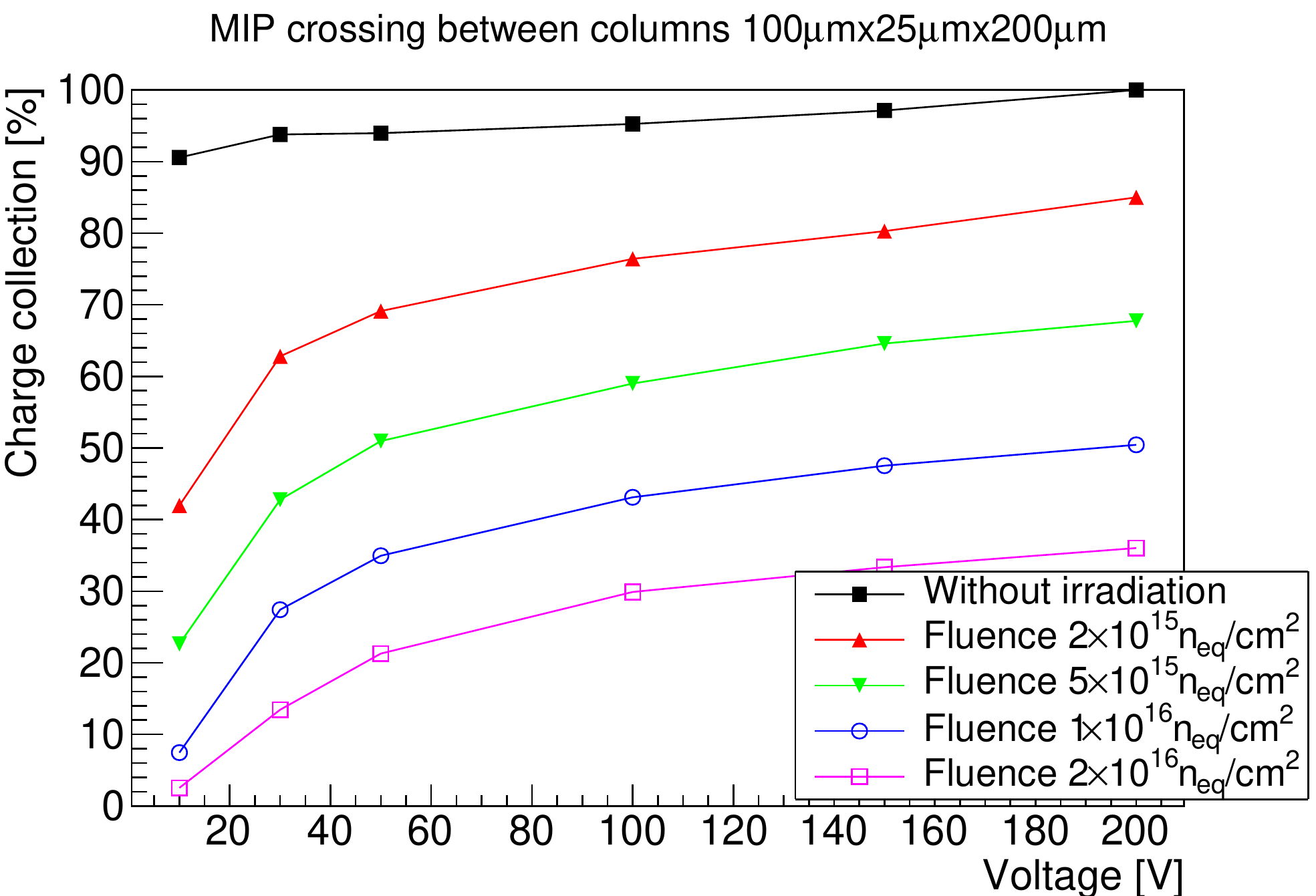}
   \caption{Collected charges for a MIP crossing between two columns of a 3D-Si \SI{100}{\micro\m}$\times$\SI{25}{\micro\m}$\times$\SI{200}{\micro\m} pixel detector.}
   \label{cc_10025}
\end{figure}

\begin{figure}[tbp!]
       \centering \includegraphics[width=\linewidth]{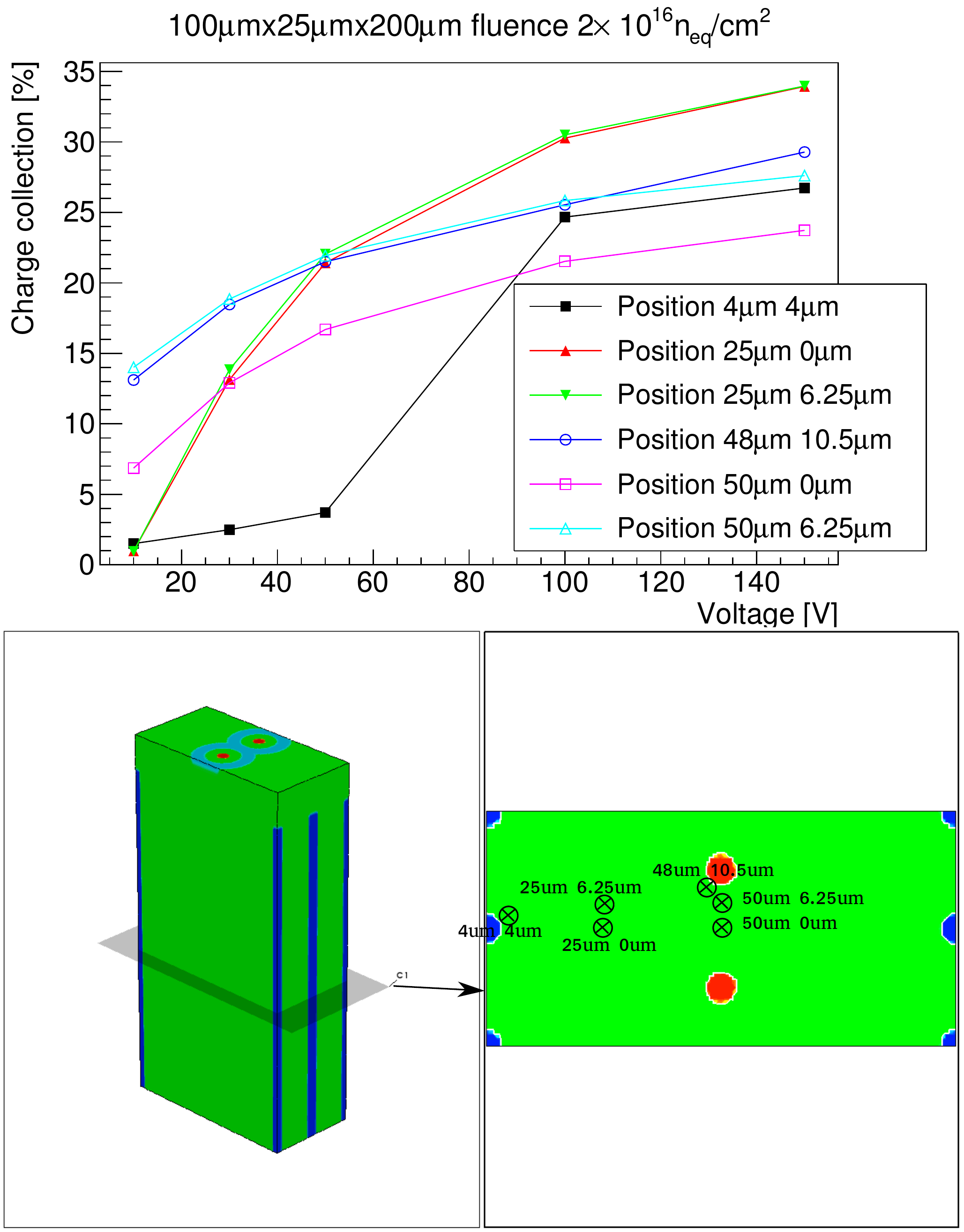}
       \caption{Collected charge for a MIP in different positions for an irradiated sensor at a fluence of \SI{2e16}{n_{eq} \cm^{-2}}. The lower part of the figure shows a cross section of the two pixel cells with the different positions of the MIP.}
       \label{c10025_f2e16}
\end{figure}

Fig. \ref{cc_10025} shows the collected charge of a MIP impinging the detector between the two columns of the pixel (position \SI{25}{\micro\m}, \SI{6.25}{\micro\m} of fig. \ref{c10025_f2e16} low figure) at different fluences. The integrated charge took into account a single pixel cell, neglecting the neighboring pixels. At fluences of \SI{2e16}{n_{eq} \cm^{-2}}, the detector shows a charge collection efficiency up to 30\% for \SI{200}{\V} of bias voltage. 
 
Fig. \ref{c10025_f2e16} shows the integrated charge of an irradiated pixel cells of \SI{100}{\micro\m}$\times$\SI{25}{\micro\m}$\times$\SI{200}{\micro\m} with a fluence of \SI{2e16}{n_{eq} \cm^{-2}}, for a MIP crossing the pixel at different positions (indicated in the low part of the figure). The charge is integrated for \SI{25}{\ns} and normalized to the non-irradiated simulation which shows a CCE close to 100\% for all the positions. The simulation used 2 pixel cells in order to take into account any collected charge by the neighboring pixels. Fig. \ref{c10025_f2e16} shows also a similar result as the ones presented in ref. \cite{Mendicino2015}.

 \section{Detectors for large $\eta$ angles}
 \label{Large_eta}
 
 3D-Si sensors located at large $\eta$ angles in the ATLAS experiment will receive particles impinging almost perpendicularly to the columns\cite{Viel2015}. Since the columns do not cross the whole volume of the detector for the IMB-CNM 3D-Si sensor design, those detectors do not show a homogeneous collected charge. Ref. \cite{Lopez2015} reports a test beam studying the collected charge of 3D-Si sensors at large $\eta$ angles and the observed inhomogeneities with the collected charge for the 3D-Si IMB-CNM design. TCAD simulations are carried out to observe the inhomogeneity of the collected charge in the detector for the FE-I4 pixel size (pixel size compatible with the FE-I4 chip size), in order to corroborate the simulations with the test beam data of ref. \cite{Lopez2015}. Since the new ATLAS upgrade will work with smaller pixel sizes, this paper shows also the simulation of the CCE for a 3D-Si \SI{50}{\micro\m}$\times$\SI{50}{\micro\m} pixel cell with the particles impinging perpendicularly to the columns in order to study its performance at large $\eta$ angles.

 \subsection{\SI{50}{\micro\m}$\times$\SI{125}{\micro\m}$\times$\SI{230}{\micro\m}, simulation of test beam results}
 \label{eta_fei4}
\begin{figure}[htb!]
\centering \includegraphics[width=\linewidth]{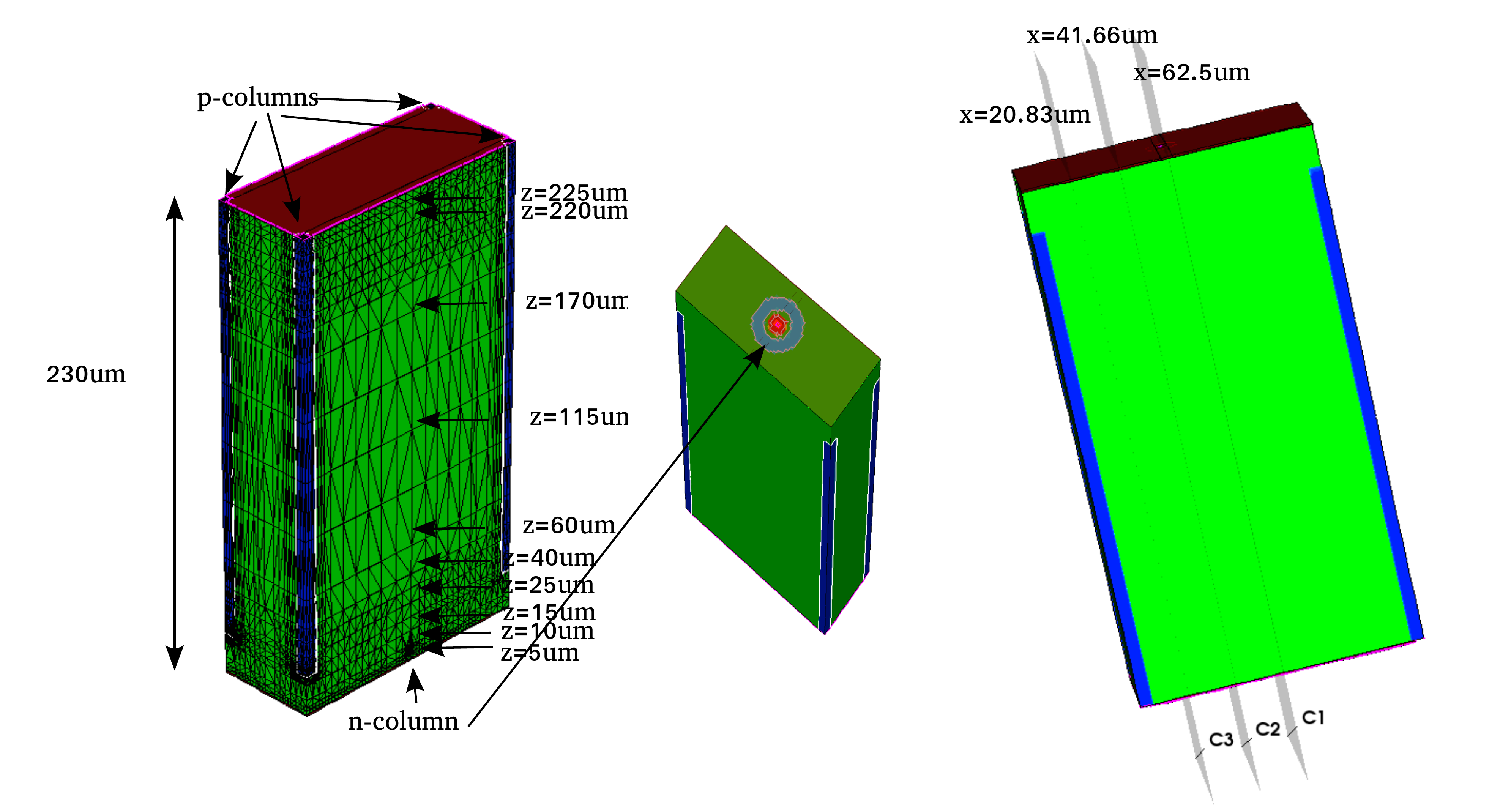}
   \caption{$z$ positions of the simulated particles (left) and $x$ planes (right).}
   \label{z_positions_largeangles}
\end{figure}

Ref. \cite{Lopez2015} reports the results of a test beam carried out with a FE-I4 detector tilted $80^\circ$ from a \SI{4}{\GeV} electron beam. 
The simulations are carried out for an FE-I4 compatible pixel cell (\SI{50}{\micro\m}$\times$\SI{125}{\micro\m}$\times$\SI{230}{\micro\m}) 3D-Si sensor, with MIP particles impinging perpendicularly to the columns perpendicularly to the \SI{125}{\micro\m} side of the detector. Since the particle might strike the pixel in different positions, the simulations are run for different $z$ positions and three different $x$ planes, shown in fig. \ref{z_positions_largeangles}. The three $x$ planes are only in one side of the detector, due to the symmetry the same results are expected for the other half. For simplification, the MIPs of the simulation impinge for a constant $z$ and $x$ position.

\begin{figure}
\centering
\begin{subfigure}{.5\linewidth}
  \centering
  \includegraphics[width=\linewidth]{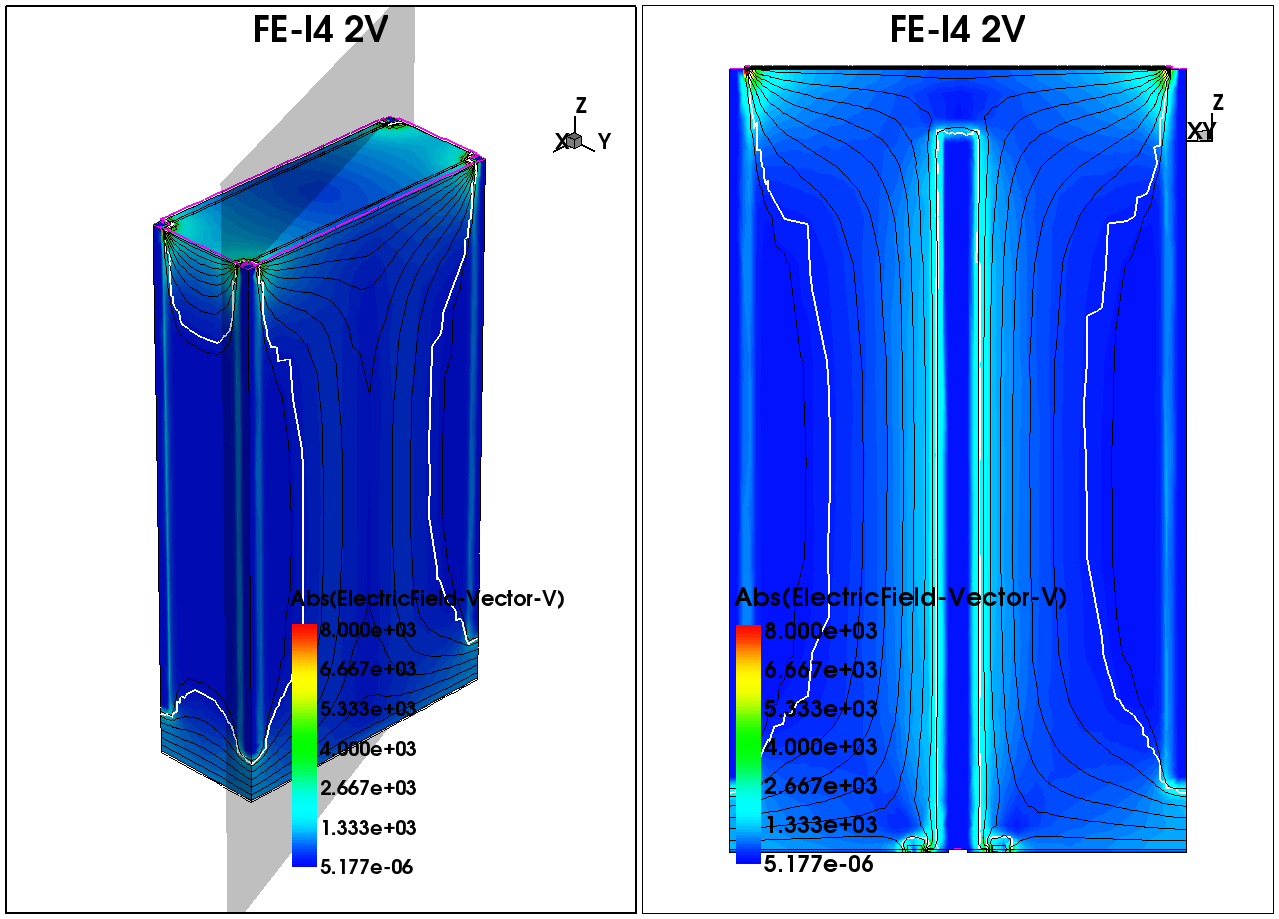}
  \caption{Bias voltage \SI{2}{\V}}
  \label{z_positions_E2V}
\end{subfigure}%
\begin{subfigure}{.5\linewidth}
  \centering
  \includegraphics[width=\linewidth]{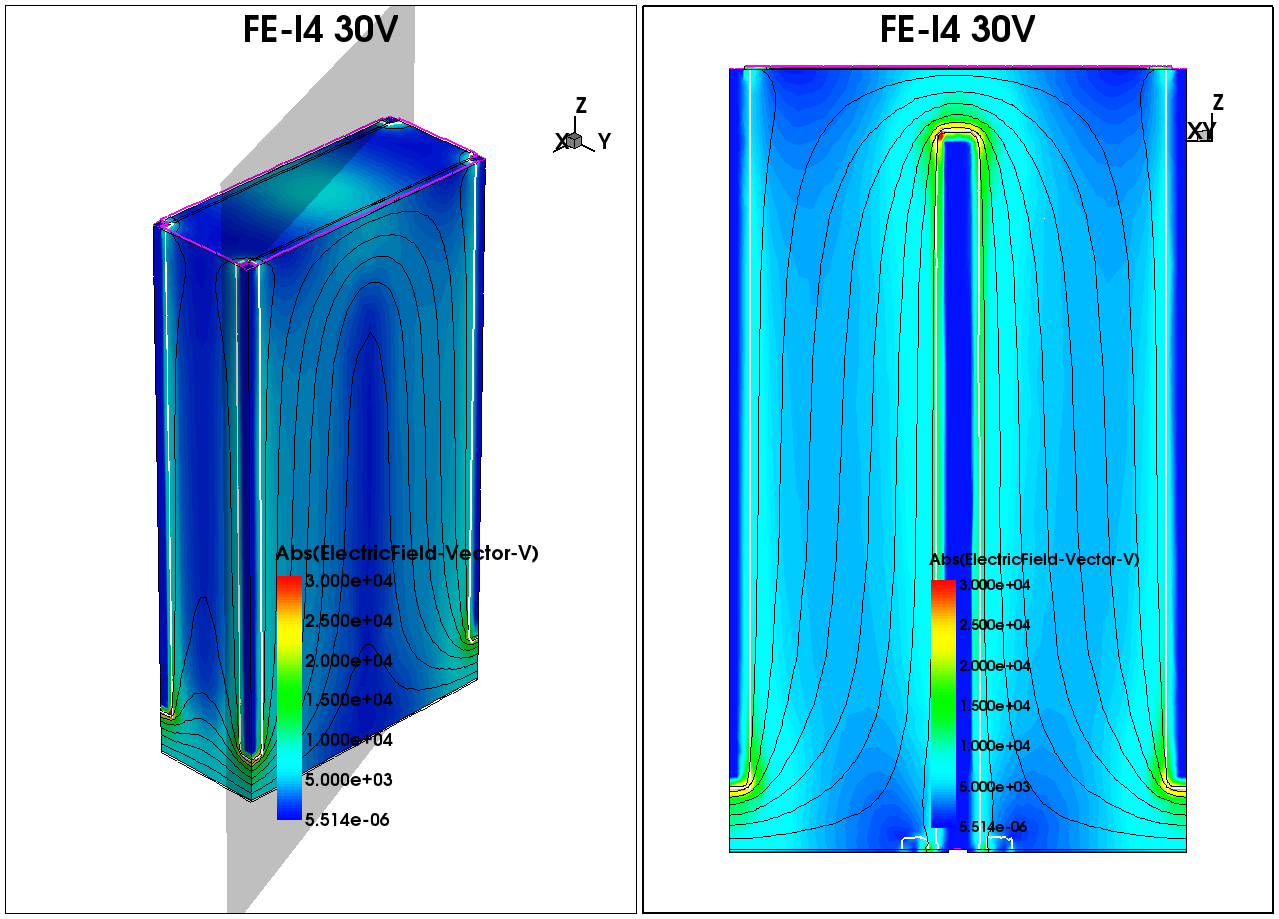}
  \caption{Bias voltage \SI{30}{\V}}
  \label{z_positions_E30V}
\end{subfigure}

\caption{Simulation of the electric field for a \SI{50}{\micro\m}$\times$\SI{125}{\micro\m}$\times$\SI{230}{\micro\m} pixel cell for two bias voltages, \SI{2}{\V} and \SI{30}{\V}. The left image is the pixel cell and the right image is the diagonal cross section of the pixel cell. The black lines are equipotential lines and the white line is the depletion region.}
\label{z_positions}
\end{figure}

The simulations are carried out for two bias voltages, \SI{2}{\V} and \SI{30}{\V}, the minimum and maximum bias voltage considered during the test beam, with the detector under depleted and fully depleted. 
Fig. \ref{z_positions}  show the simulation of the electric field in a diagonal cut that includes two p-columns and the n-column for \SI{2}{\V} and \SI{30}{\V}, respectively. The white line of the simulation corresponds to the depleted volume, and the simulation for \SI{2}{\V} shows that the pixel volume is not fully depleted. The electric field simulations (figs. \ref{z_positions_E2V} and \ref{z_positions_E30V}) show that the electric field at the tip of the n-column changes from the under depleted to the depleted detector, showing a negative electric field gradient at high voltages from n to p-column, whereas at \SI{2}{\V} shows a positive electric field gradient thus the collected charge has a negative polarization, and a negative CCE.

\begin{figure}[htb!]
\centering \includegraphics[width=\linewidth]{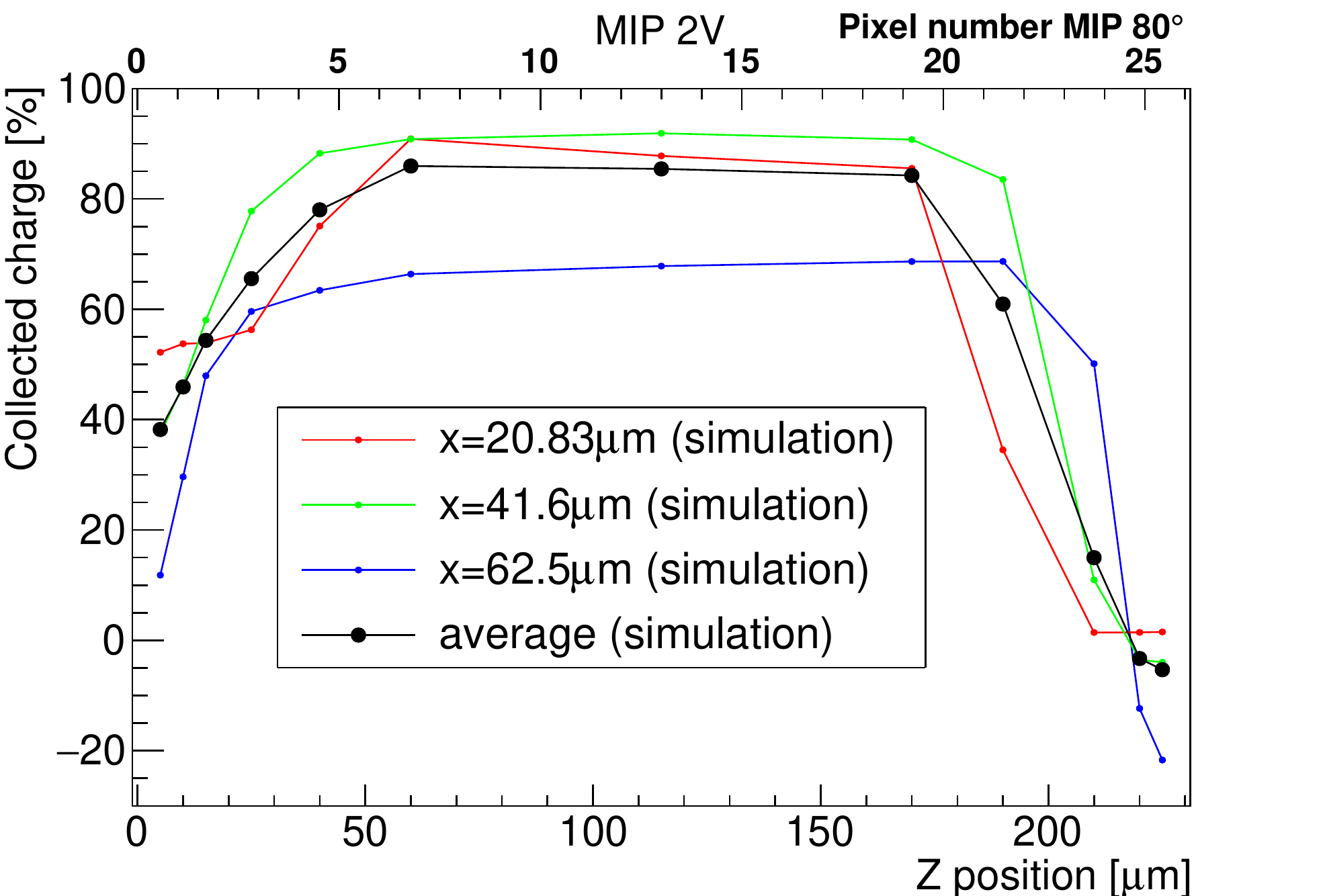}
   \caption{Simulation of MIP particles through different positions of the \SI{50}{\micro\m}$\times$\SI{125}{\micro\m}$\times$\SI{230}{\micro\m} pixel cell at \SI{2}{\V}. The upper axis corresponds to the pixel number of the experiment for large $\eta$, tilted $80^\circ$ from the beam.}
   \label{MIP2_largeangles}
\end{figure}

Fig. \ref{MIP2_largeangles} shows the simulation of the CCE generated by a MIP particle crossing different positions of the FE-I4 geometry pixel cell detector at \SI{2}{\V}. 
The simulations results are biased at \SI{2}{\V} since the testbeam data showed a decrement of the active volume at low bias voltages. 
The average charge is weighted with half the charge for $x=$\SI{62.5}{\micro\m} position, since the position belonging to the n-column is less probable than the others:

\begin{equation}
 Q_{average}= \frac{2Q_{x20.83}+2Q_{x41.6}+Q_{x62.5}}{5}
 \label{average_charge}
\end{equation}

where $Q_{x20.83} $ is the integrated charge of the position x=\SI{20.83}{\micro\m}, $Q_{x41.6} $ is the integrated charge of the position x=\SI{41.6}{\micro\m} and $Q_{x62.5}$ is the integrated charge of the position x=\SI{62.5}{\micro\m}. 
The MIPs crossing the n-column is less probable than the other positions because in the whole pixel cell there is one n-column but the positions $x=$\SI{41.66}{\micro\m} and $x=$\SI{20.83}{\micro\m} are in both sides of the n-column, for that reason the charge in the other two positions are weighted two times more than the charge at the n-column position.

\begin{figure}[htb!]
\centering \includegraphics[width=\linewidth]{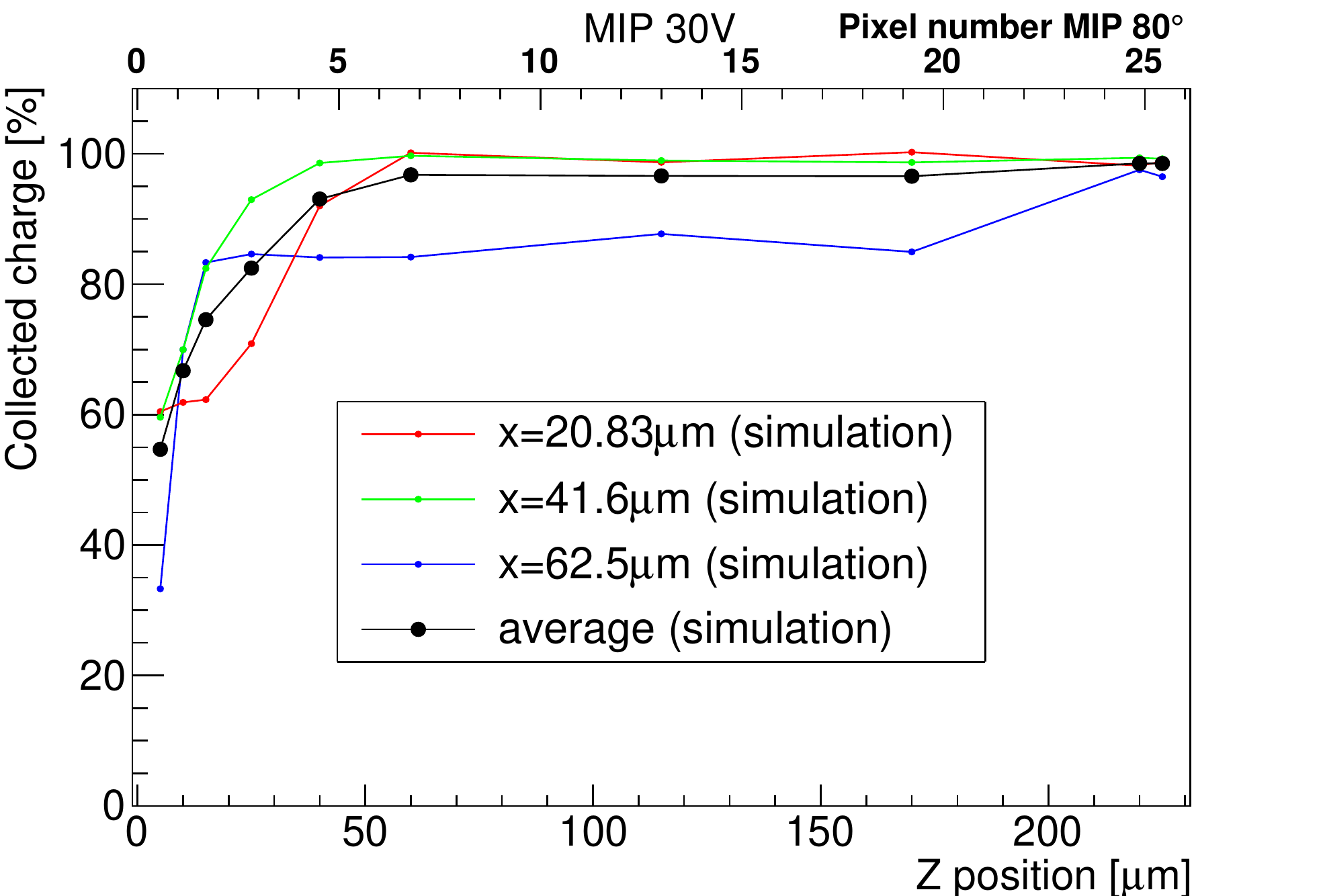}
   \caption{Simulation of MIP particles through different positions of the \SI{50}{\micro\m}$\times$\SI{125}{\micro\m}$\times$\SI{230}{\micro\m} pixel cell at \SI{30}{\V}. The upper axis corresponds to the pixel number of the experiment for large $\eta$, tilted $80^\circ$ from the beam.} 
   \label{MIP30_largeangles}
\end{figure}

The simulation for \SI{2}{\V} shows a similar behaviour as the one reported in ref. \cite{Lopez2015}, and the decrement of the active volume correspond to the changes of the electric field for small $z$ positions at low voltages (as shown in fig. \ref{z_positions_E2V}). Fig. \ref{MIP30_largeangles} shows the simulation of a MIP passing through the detector at different positions with a bias voltage of \SI{30}{\V}. The average charge is calculated with the same weighting charge as the one in equation \ref{average_charge}. The simulations at \SI{30}{\V} show good agreement with the testbeam data\cite{Lopez2015}, since the testbeam data shows a low signal for low $z$ for the IMB-CNM 3D-Si sensor as reported in the simulations (for low and high bias voltages).

\subsection{Detectors with a \SI{50}{\micro\m}$\times$\SI{50}{\micro\m} pixel cell}

A similar simulation was carried out for 3D-Si sensor with a pixel cell of \SI{50}{\micro\m}$\times$\SI{50}{\micro\m}$\times$\SI{200}{\micro\m} (the same used in section \ref{mip5050}), in order to study the behaviour of the detectors at large $\eta$ angles for the new ATLAS geometry. Fig. \ref{dos_cols} (left) shows the simulated pixel cell. The simulation took into account two bias voltages, \SI{2}{\V} and \SI{30}{\V} as the ones simulated in section \ref{eta_fei4} in order to compare with the results shown in the previous section.

\begin{figure}
\centering
\begin{subfigure}{.5\linewidth}
  \centering
  \includegraphics[width=\linewidth]{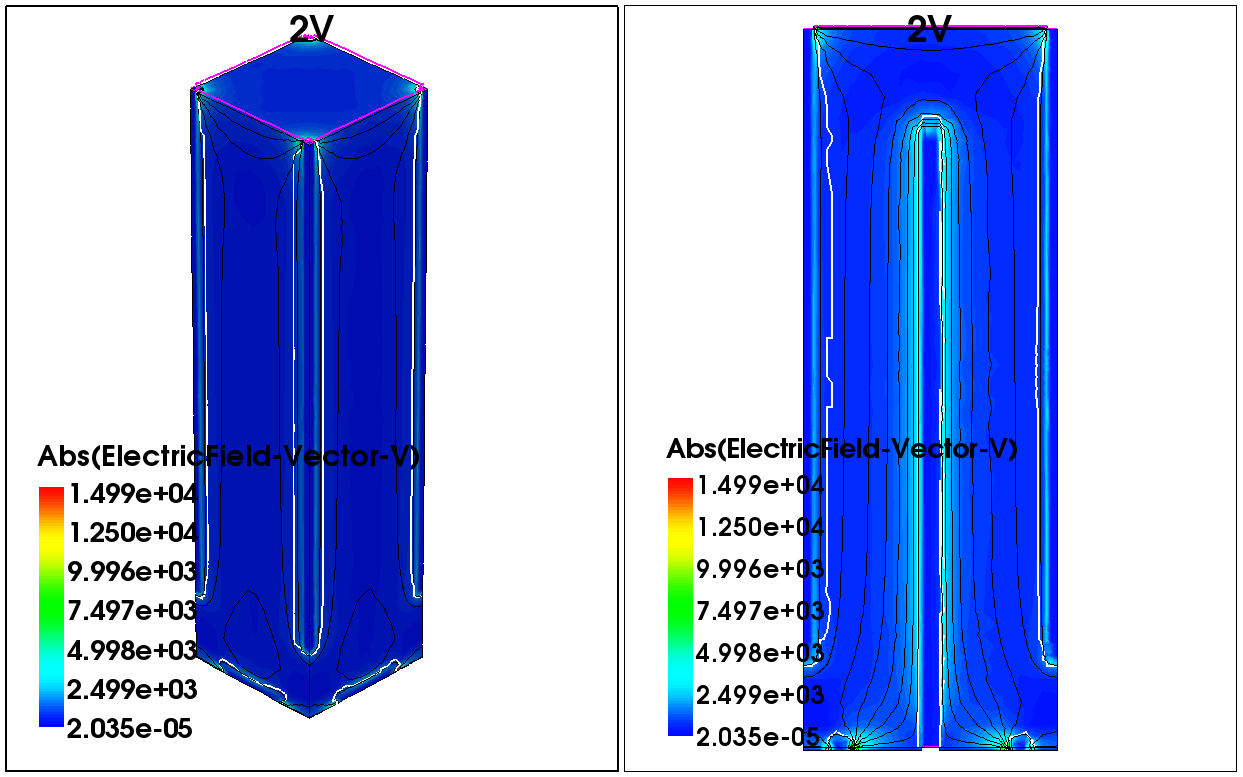}
  \caption{Bias voltage \SI{2}{\V}}
  \label{c5050_E2V}
\end{subfigure}%
\begin{subfigure}{.5\linewidth}
  \centering
  \includegraphics[width=\linewidth]{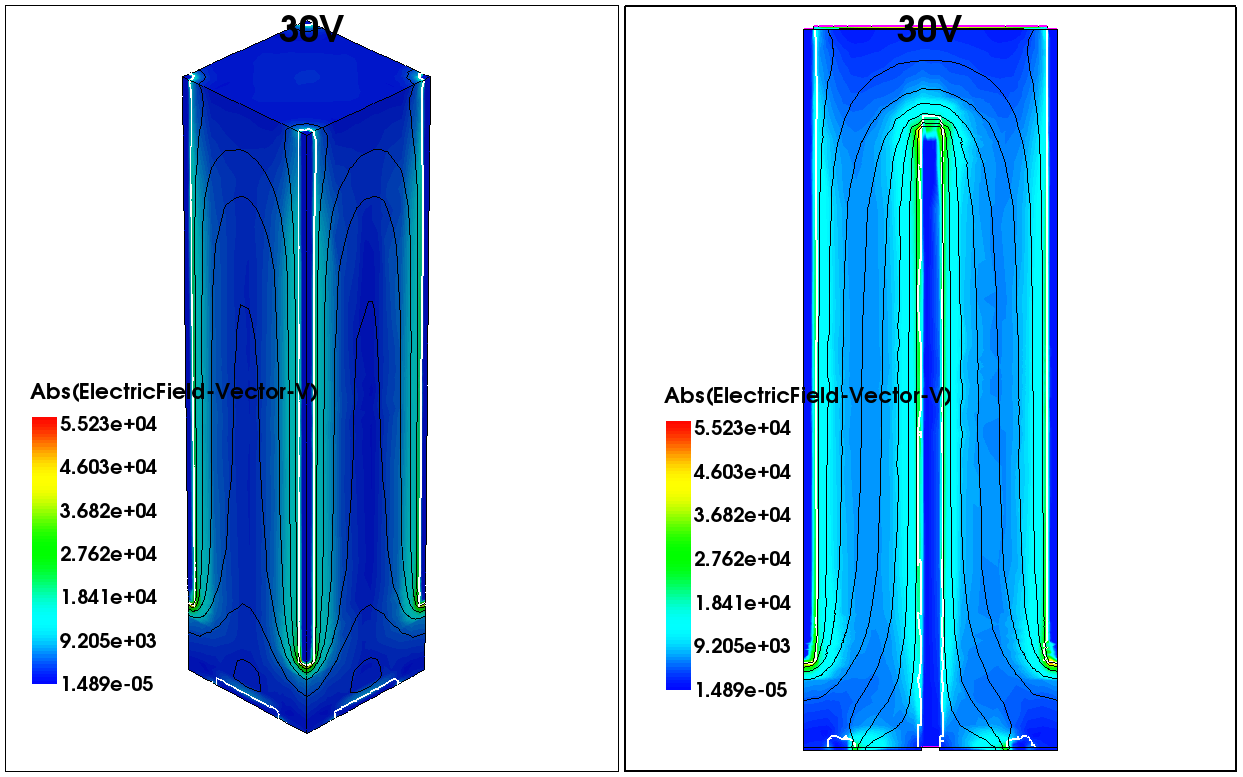}
  \caption{Bias voltage \SI{30}{\V}}
  \label{c5050_E30V}
\end{subfigure}
\caption{Simulation of the electric field for a \SI{50}{\micro\m}$\times$\SI{50}{\micro\m}$\times$\SI{200}{\micro\m} for two bias voltages, \SI{2}{\V} and \SI{30}{\V}. The left image is the pixel cell and the right image is the diagonal cross section of the pixel cell. The black lines are equipotential lines and the white line is the depletion region.}
\label{electric_field_5050}
\end{figure}

Figs. \ref{c5050_E2V} and \ref{c5050_E30V} show the simulation of the electric field for \SI{2}{\V} and \SI{30}{\V}, respectively. The pixel shows almost full depletion at \SI{2}{\V}. 
The simulations are carried out for the same bias voltage as in fig. \ref{z_positions} in order to compare the results.

%

\begin{figure}[htb!]
\centering \includegraphics[width=\linewidth]{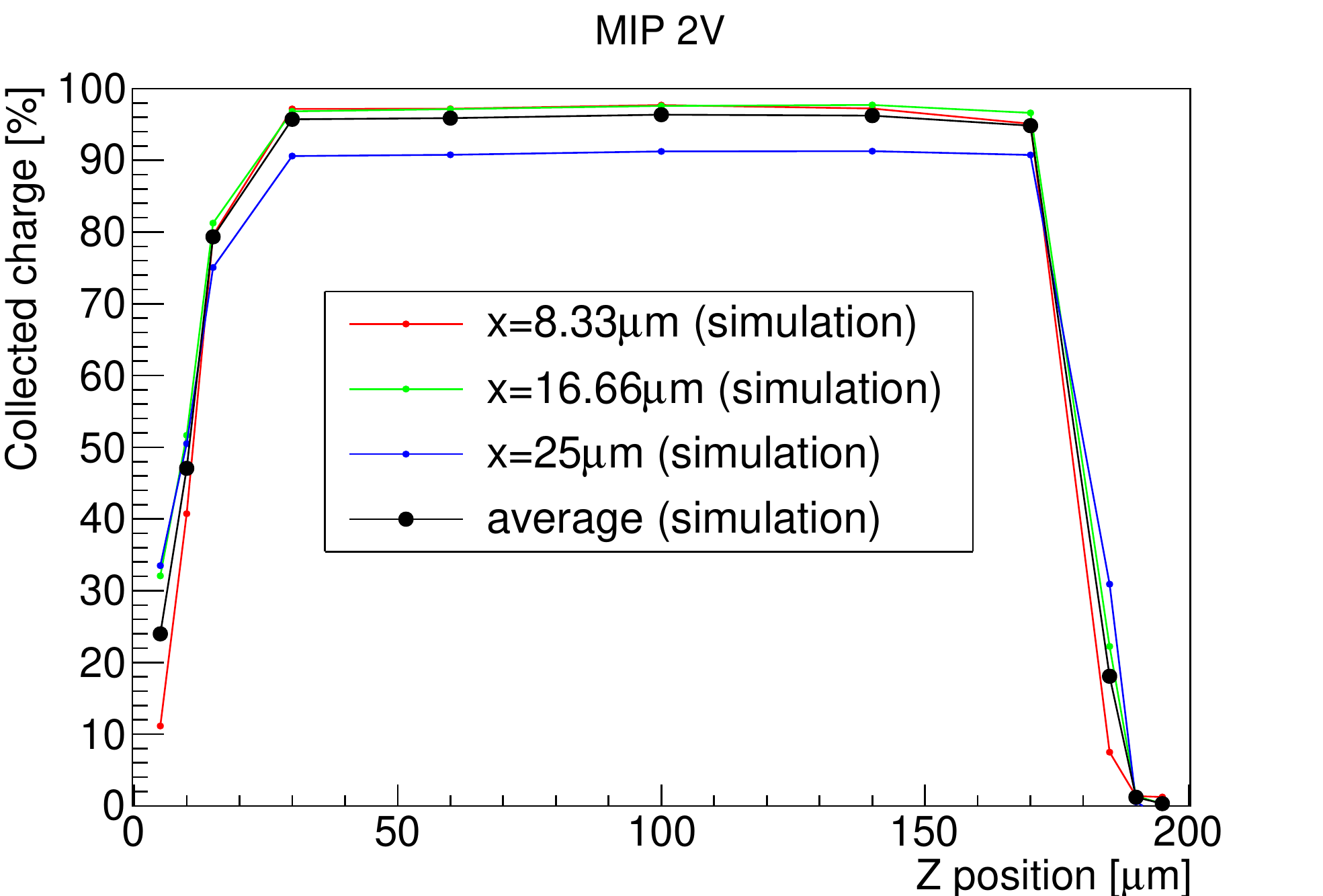}
   \caption{Simulation of a MIP impinging perpendicularly to the columns of a \SI{50}{\micro\m}$\times$\SI{50}{\micro\m}$\times$\SI{200}{\micro\m} detector biased at \SI{2}{\V}.}
   \label{cc5050_2V}
\end{figure}

\begin{figure}[htb!]
       \centering \includegraphics[width=1\linewidth]{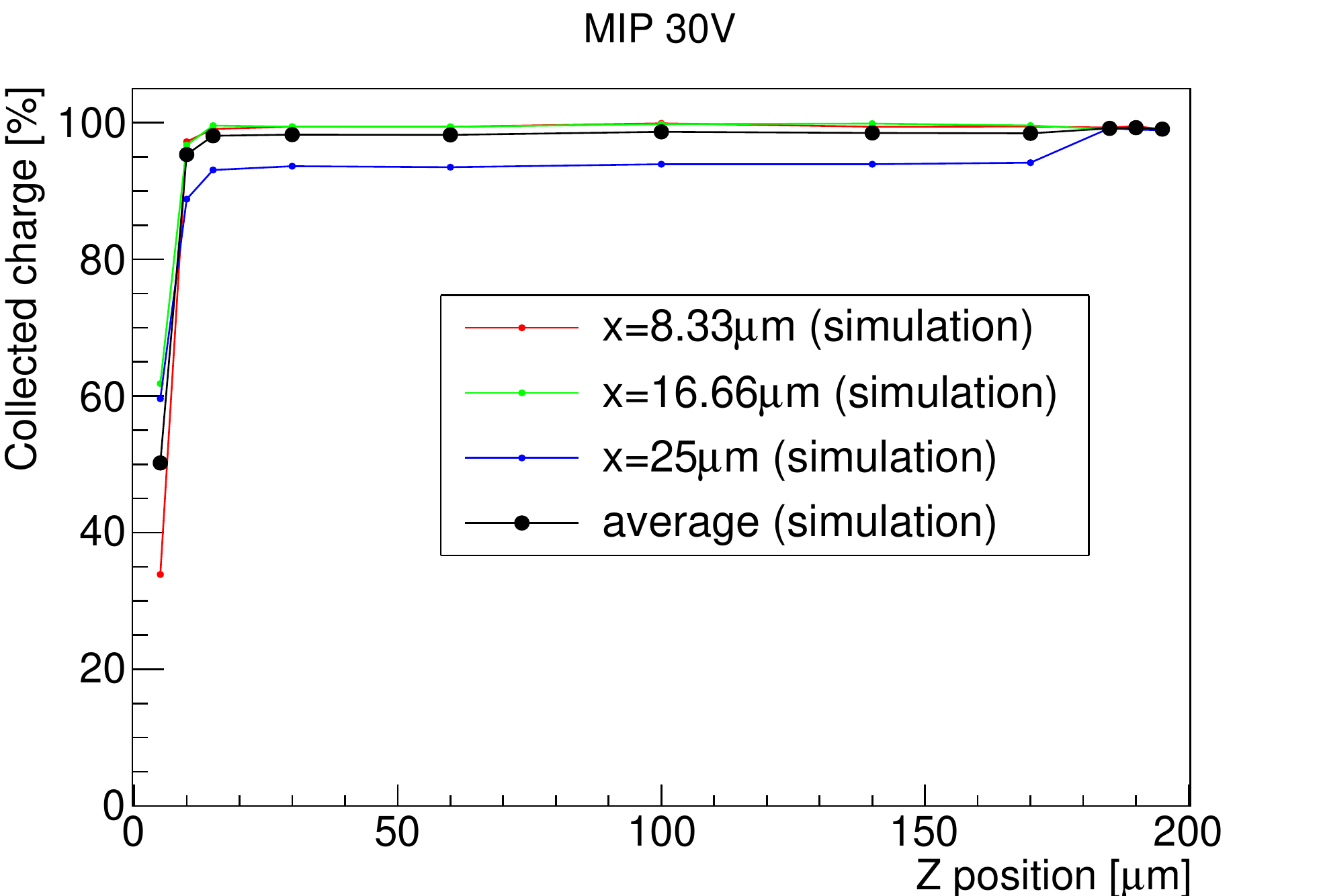}
   \caption{Simulation of a MIP impinging perpendicularly to the columns of a \SI{50}{\micro\m}$\times$\SI{50}{\micro\m}$\times$\SI{200}{\micro\m} detector biased at \SI{30}{\V}.}
       \label{cc5050_30V}
\end{figure}

Figs. \ref{cc5050_2V} and \ref{cc5050_30V} show the simulation of the collected charge generated by a MIP crossing different position, in similar locations as the ones detailed in fig. \ref{MIP2_largeangles} but for the \SI{50}{\micro\m}$\times$\SI{50}{\micro\m}$\times$\SI{200}{\micro\m} for \SI{2}{\V} and \SI{30}{\V} bias voltages, respectively. The $x$ planes considered are $x=$\SI{8.33}{\micro\m}, $x=$\SI{16.66}{\micro\m} and $x=$\SI{25}{\micro\m} that correspond to 1/3 of the distance between the edge and the center of the column, 2/3 and the plane that crosses the n+ column, respectively, only taking into account half of the geometry for symmetry. The average is calculated as:

\begin{equation}
 Q_{average}= \frac{2Q_{x8.33}+2Q_{x16.66}+Q_{x25}}{5}
 \label{average_charge5050}
\end{equation}
where $Q_{x8.33}$ is the integrated charge of the position x=\SI{8.33}{\micro\m}, $Q_{x16.66}$ is the integrated charge of the position x=\SI{16.66}{\micro\m} and $Q_{x25}$ is the integrated charge of the position x=\SI{25}{\micro\m}. As before, the charge belonging to the column position is weighted half the charge of the other positions since it is less probable.


Fig. \ref{cc5050_2V} does not show a negative signal at large $z$ positions as the simulation for the \SI{50}{\micro\m}$\times$\SI{125}{\micro\m}$\times$\SI{230}{\micro\m} in fig. \ref{MIP2_largeangles}, showing more active area than the simulation of \SI{50}{\micro\m}$\times$\SI{125}{\micro\m}$\times$\SI{230}{\micro\m} pixel cell. The collected charge at \SI{30}{\V} shows very good performance since at $z=$\SI{10}{\micro\m} the CCE almost reach 100\%.

\section{3D-Si single sided sensors}

Future ATLAS and LHC specifications for pixel detectors might need thinner detectors. Detectors located at large $\eta$ angles, far from the interaction point, will receive the particles almost perpendicularly to the column direction. In order to have a more precise detection at those positions, thinner detectors are proposed\cite{Viel2015}. 

\begin{figure} 
\centering \includegraphics[width=\linewidth]{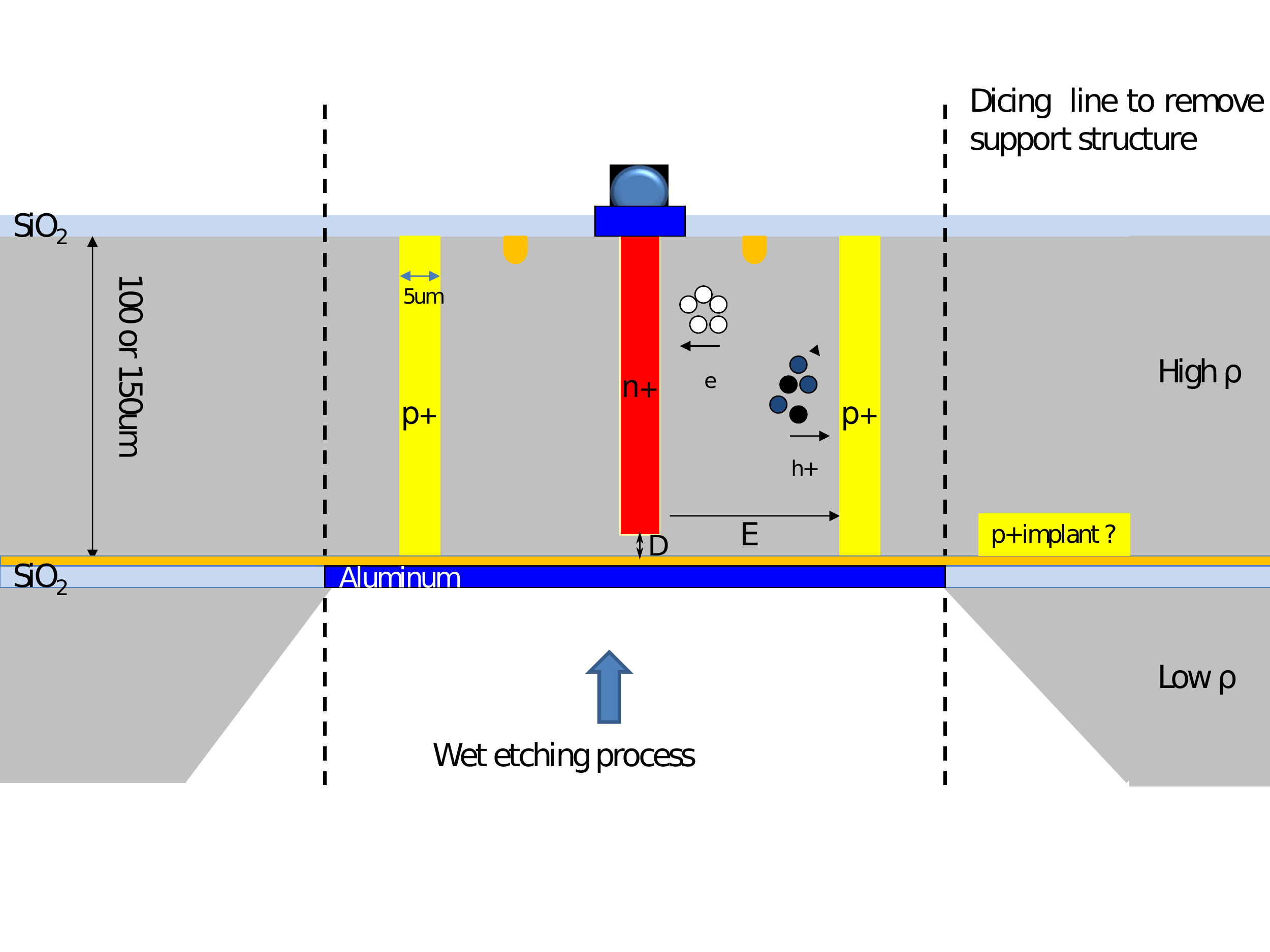}
   \caption{Scheme for 3D-Si single side detector.}
   \label{3D_single_side_schematic}
\end{figure}

Due to the complex fabrication process of the 3D-Si sensors, the best solution to fabricate thin 3D-Si sensors is on a Silicon-On-Oxide (SOI) wafer. A high resistivity thin silicon wafer is separated from a thicker low resistivity substrate by a Buried OXide (BOX) layer that can be removed with a wet etching process. IMB-CNM used this technique with several fabrications as the ones reported in ref. \cite{Fleta2014, Fleta2015a}. Fig. \ref{3D_single_side_schematic} shows a schematic of the 3D-Si single sided detector proposed for the HL-LHC ATLAS experiment. 

In order to test different configurations, two possible types of wafers are proposed: a \SI{150}{\micro\m} SOI p-type wafer and \SI{150}{\micro\m} SOI p-type wafer with a p-type backside implant. The thickness of the active wafer can be reduced according to the requirement of the experiment.

\begin{figure} [tb!]
\centering \includegraphics[width=\linewidth]{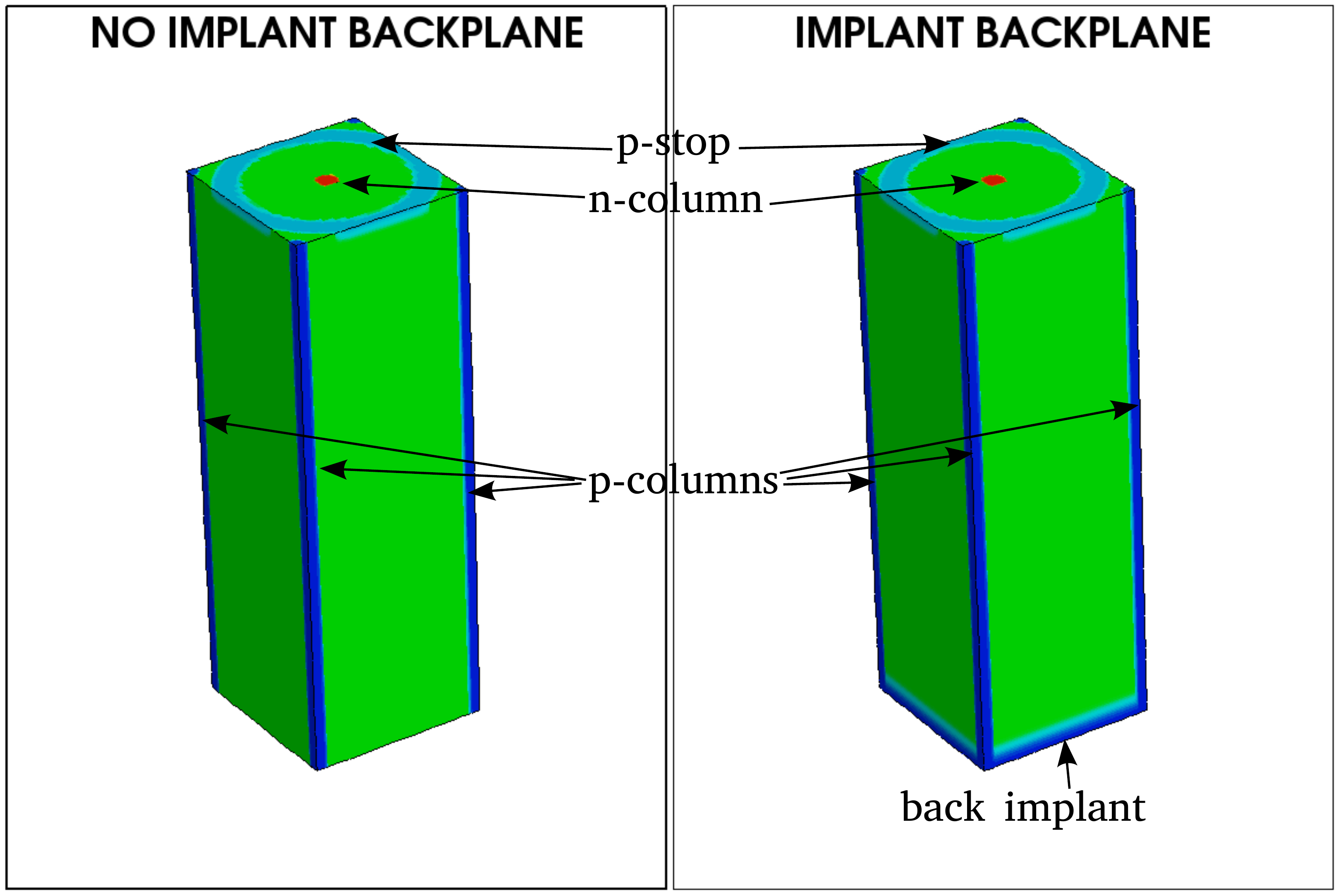}
   \caption{Net doping profile of the 3D-Si single sided without implant on the backside (left) and with implant on the backside (right).}
   \label{3Dss_implant_noimplant}
\end{figure}

Fig. \ref{3Dss_implant_noimplant} shows the simulated structures without the backside implant and with the backside implant. Both structures were simulated in order to optimize the n-column depth of these 3D-Si sensors. The simulated detectors are single sided p-type detectors with a pixel geometry of \SI{50}{\micro\m}$\times$\SI{50}{\micro\m}$\times$\SI{150}{\micro\m}. The p-columns have a depth of \SI{150}{\micro\m}, reaching the BOX, while the n-columns will be a non passing-through columns. If the n-column reach the BOX it will lead to an early break down due to the superficial currents through the silicon dioxide positive charge, unless you isolate with a p-spray implant to isolate the backplane surface. The diameter of the columns are expected to be \SI{5}{\micro\m} and the p-stop will have a radius of \SI{25}{\micro\m}.
The simulations are performed for different n-column depths, being $D$ the distance of the n-column to the BOX (depicted under the n-column in fig. \ref{3D_single_side_schematic}), beginning from a reference distance $d=$ \SI{35.35}{\micro \m}. The simulated $D$ values are $d$-\SI{10}{\micro\m}, $d$-\SI{5}{\micro\m}, $d$, $d$+\SI{5}{\micro\m} and $d$+\SI{10}{\micro\m}. The simulation of those new structures is important because the DRIE process can be controlled by a precision of $\pm$\SI{10}{\micro\m}. 

\begin{figure}
 \centering
\includegraphics[width=\linewidth]{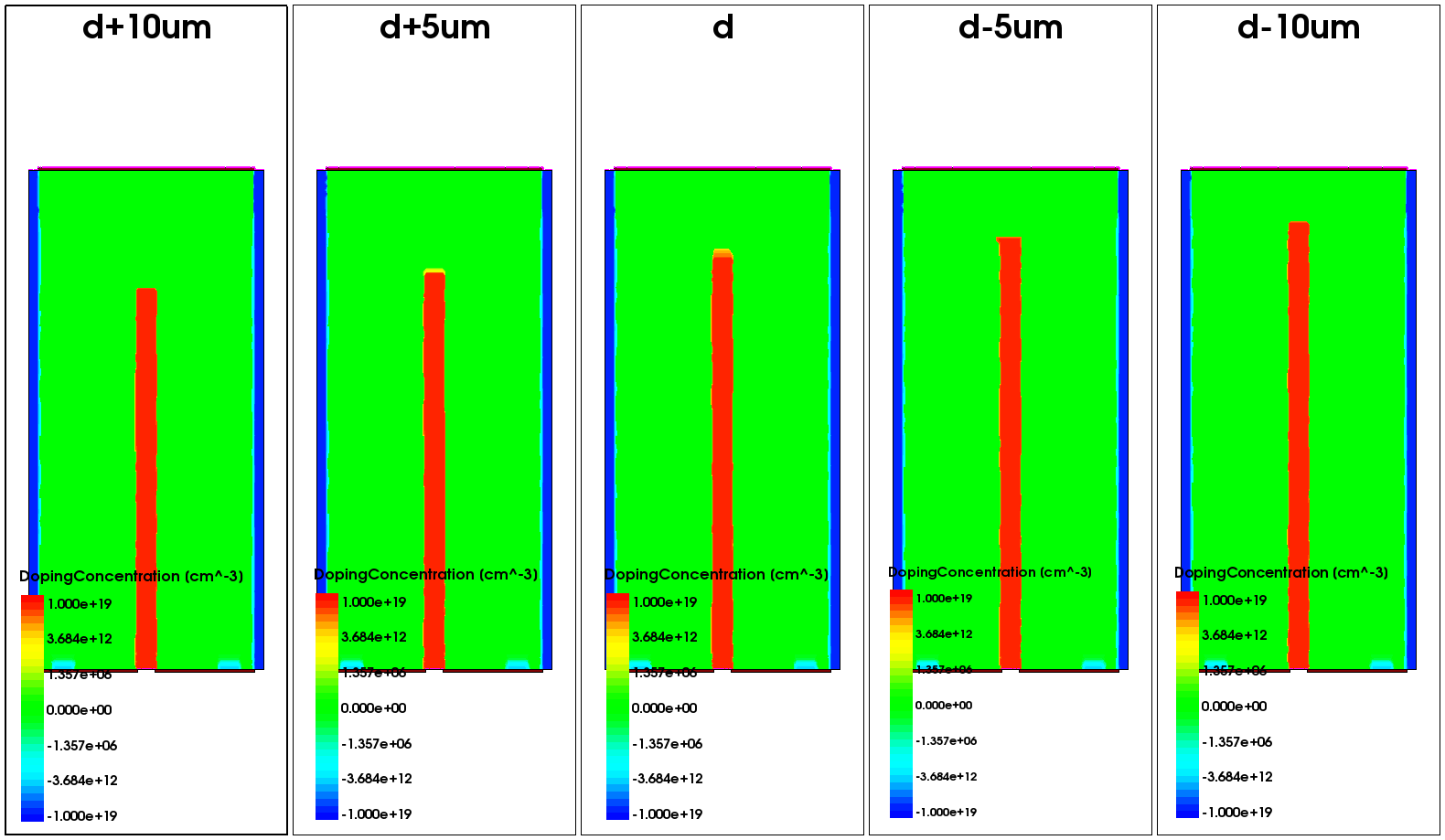} 
 \caption{Cross sections of the simulation for the 3D-Si single sided detectors with different depths of the columns without implant in the backplane. The electric field is at break-down voltage (circa \SI{140}{\V}), shown in fig. \ref{3D_SS_iv_wo}. } 
 \label{3D_wo_implant_bp}
 \end{figure}

\begin{figure}[htb!]
 \centering
 \includegraphics[width=\linewidth]{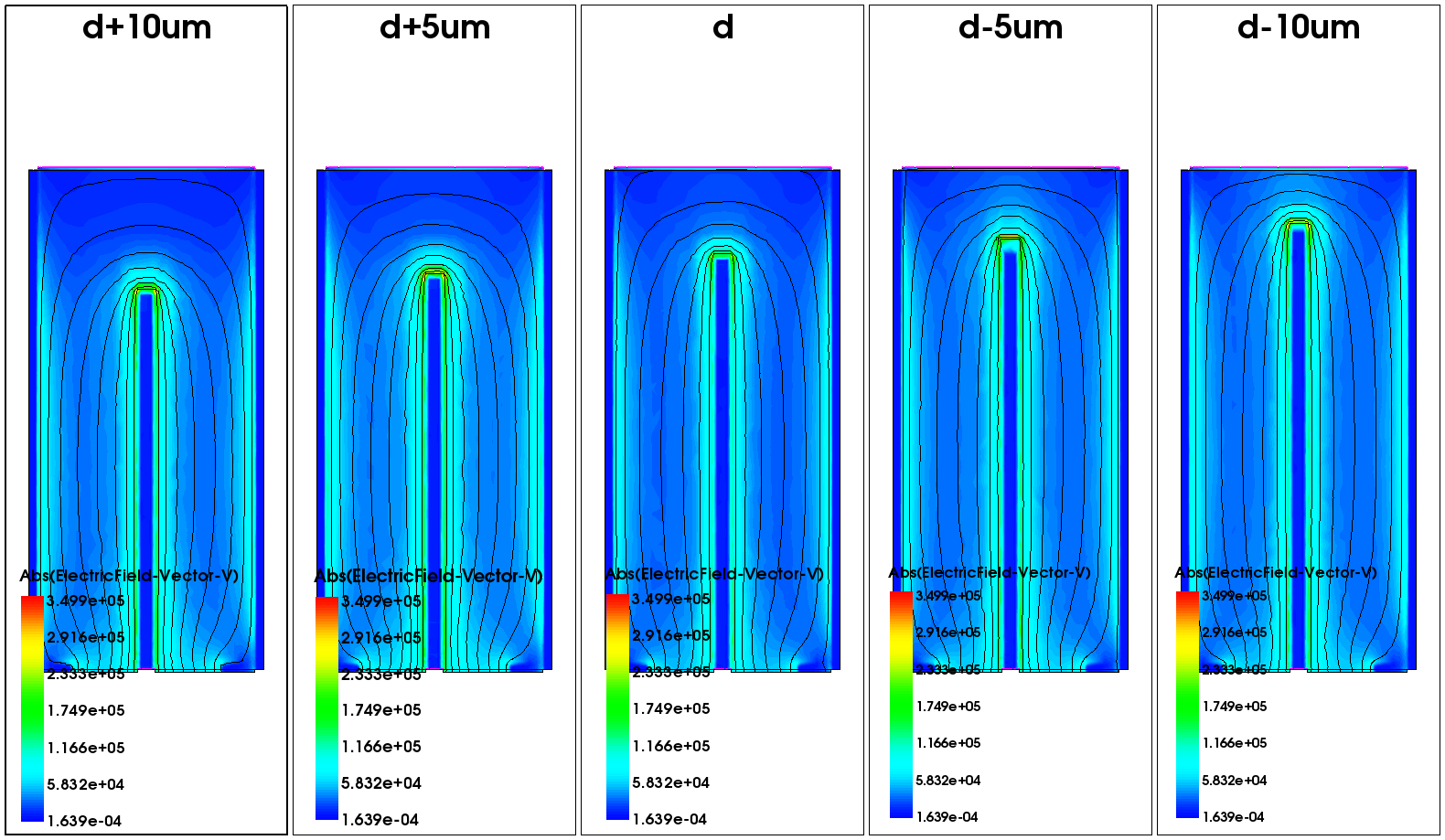}
 \caption{Cross sections of the electric field simulation for the 3D-Si single sided detectors with different depths of the columns without implant in the backplane. The black lines are equipotential lines.}
 \label{3D_wo_implant_bp_E}
 \end{figure}

 Fig. \ref{3D_wo_implant_bp} shows cross sections of the doping profile in a diagonal cut that includes the 2 p-columns and the n-column of the simulations of a detector without p-implant on the backside, and fig. \ref{3D_wo_implant_bp_E} shows their electric field.

\begin{figure}
 \centering
 \includegraphics[width=\linewidth]{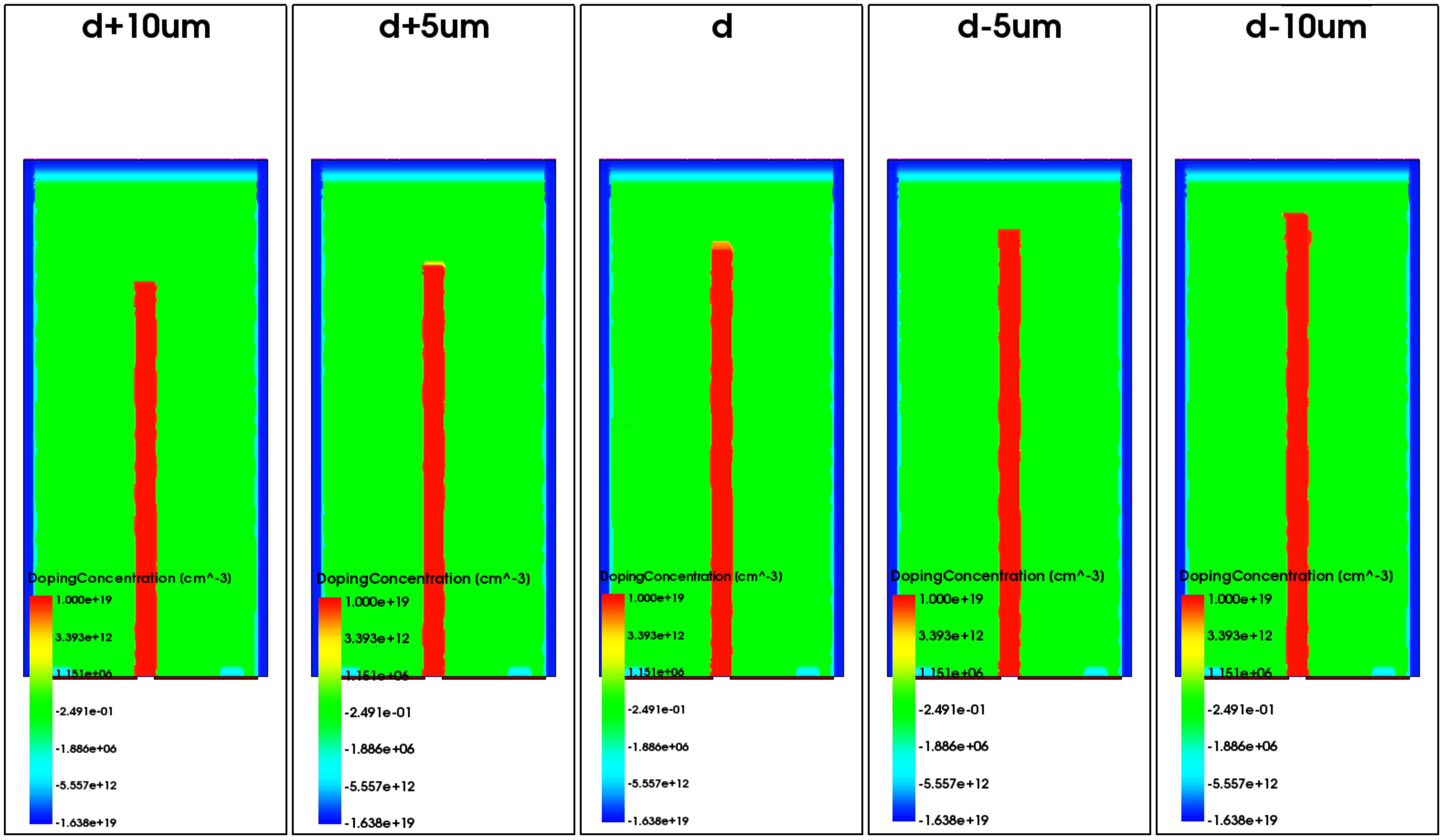}
 \caption{Cross sections of the simulation for the 3D-Si single sided detectors with different depths of the columns with implant in the backplane.}
 \label{3D_w_implant_bp}
 \end{figure}

\begin{figure}
 \centering
 \includegraphics[width=\linewidth]{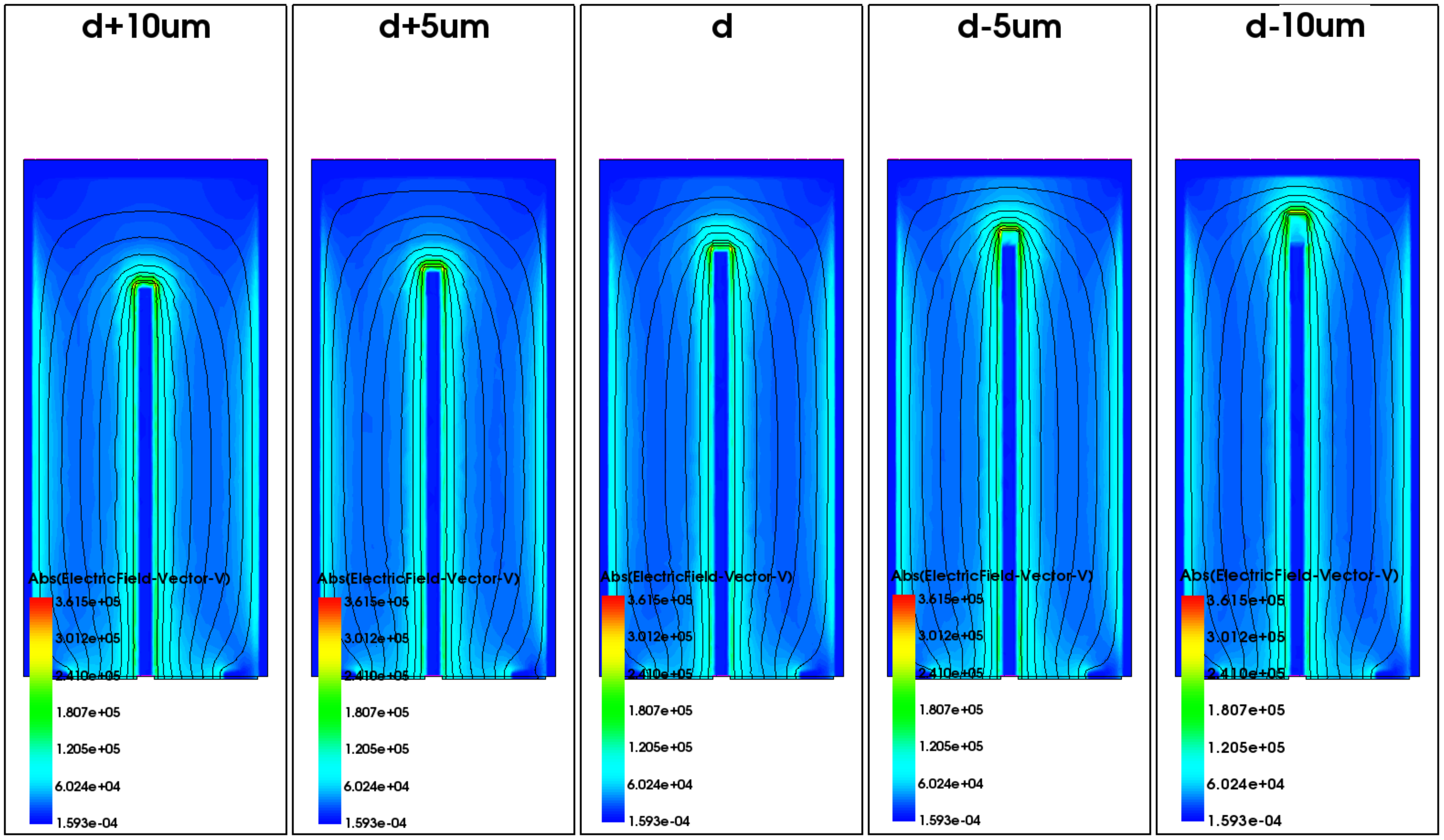}
 \caption{Cross sections of the electric field simulation for the 3D-Si single sided detectors with different depths of the columns with implant in the backplane. The electric field is at break-down voltage (circa \SI{140}{\V}), shown in fig. \ref{3D_SS_iv_w}. The black lines are the equipotential lines.}
 \label{3D_w_implant_bp_E}
 \end{figure}

 Fig. \ref{3D_w_implant_bp} shows the same cross sections as in fig. \ref{3D_wo_implant_bp} but with a p-implant on the backside. Fig. \ref{3D_w_implant_bp_E} shows the electric field with the p-implant on the backside for the different distances $D$.


\begin{figure}[htb!]
\centering{\includegraphics[width=\linewidth]{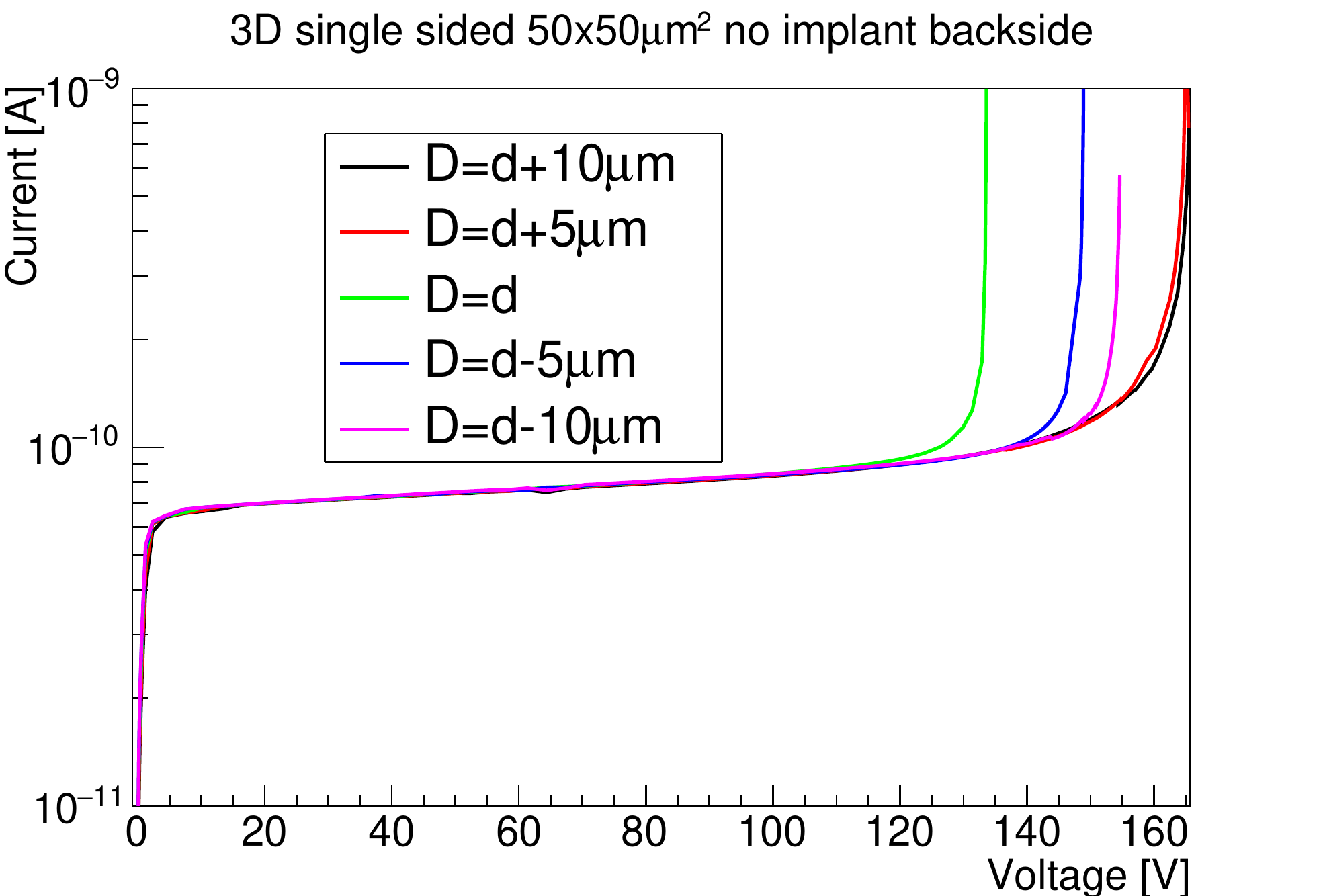}} 
\caption{Simulation of current-voltage curves for 3D single sided \SI{50}{\micro\m}$\times$\SI{50}{\micro\m}$\times$\SI{150}{\micro\m} detector.}
\label{3D_SS_iv_wo}
\end{figure}

\begin{figure}[htb!]
\centering\includegraphics[width=\linewidth]{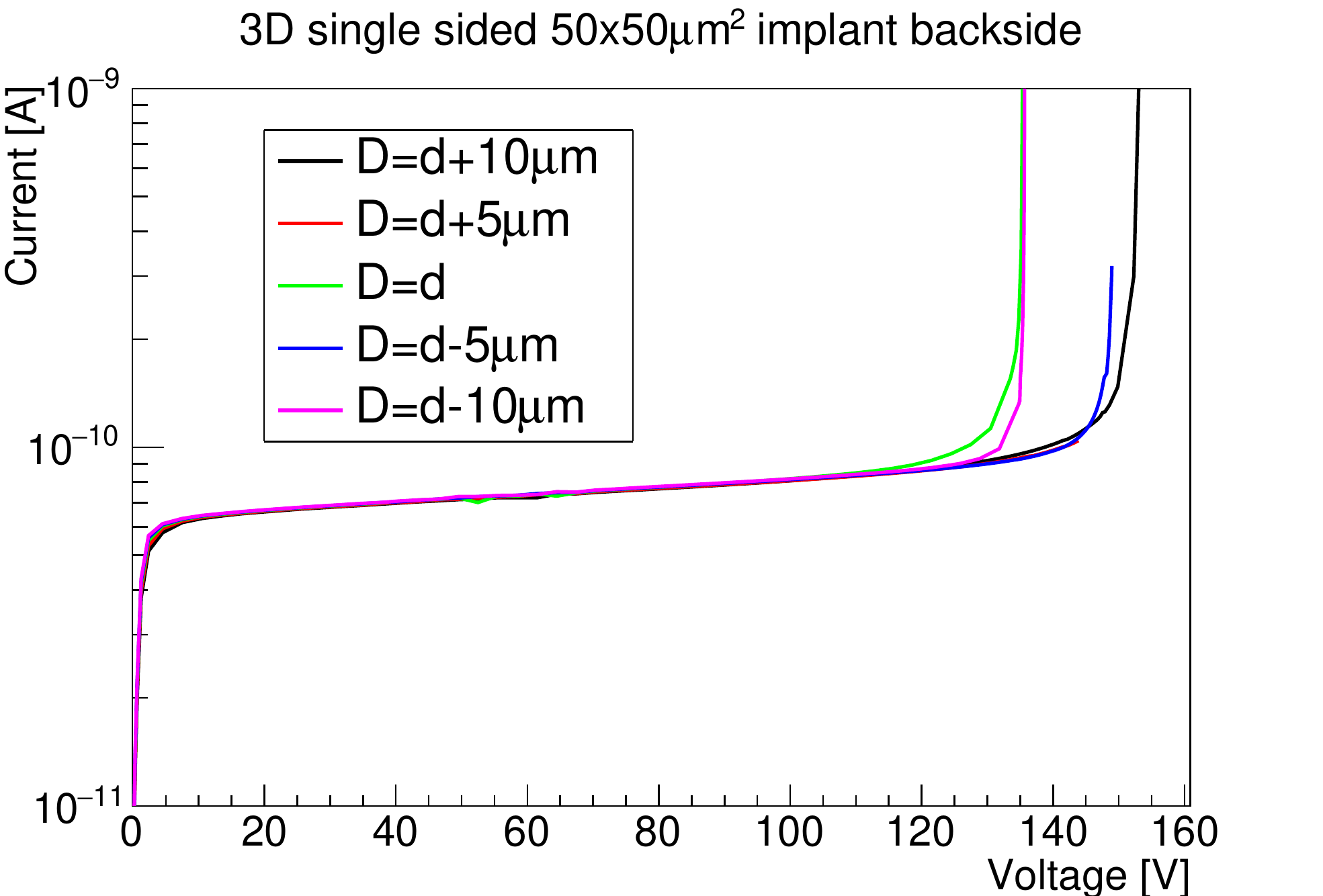}
\caption{Simulation of current-voltage curves for 3D single sided \SI{50}{\micro\m}$\times$\SI{50}{\micro\m}$\times$\SI{150}{\micro\m} detector.}
\label{3D_SS_iv_w}
\end{figure}

Fig. \ref{3D_SS_iv_wo} shows the current-voltage curves for the 3D-Si single sided detectors without p-implant in the backside and fig. \ref{3D_SS_iv_w} shows the current-voltage curves of the detectors with the p-implant at the backside.

 The detector without the p-implant in the backside has slightly larger break down voltages than the detectors with the p-implant. Besides that, there is not any relevant difference in the leakage currents between the pixel with p-implant in the backside and without it, as it is also reported in the ref. \cite{Pennicard2010a}. Anyway, the presence of the p-implant on the backside will simplify the contact of the aluminum layer deposited on the back surface of the wafer to apply the high bias voltage. In the case of the wafer without p-implant the p$^+$ holes have to be etched through the BOX oxide in order to reach the aluminum layer used to bias the detector from the back surface.

 \section{Conclusions}
 
 The simulations of new geometries with smaller pixel sizes show good response without irradiation and their collected charge will decrease to less than half when they are irradiated with the expected fluences of \SI{2e16}{n_{eq}\per \cm^2}.
 
  3D-Si sensors located at large $\eta$ angles will have particles impinging almost perpendicularly to the columns and data collected during a test beam showed that IMB-CNM 3D-Si sensors do not have an homogeneous charge collection due to the non passing through columns. The simulations carried out to study this effect show good agreement with the test beam data. The simulations for the new geometry (\SI{50}{\micro\m}$\times$\SI{50}{\micro\m}) show almost all active volume of CCE for large $\eta$ positions.
 
 Thinner detectors fabricated on SOI wafers are studied for large $\eta$ angles. According to the simulations, the presence of a p-implant on the backside will not change electrically the behaviour of those sensors but should simplify the fabrication process. The expected CCE is the same as the simulations of the double sided with columns \SI{50}{\micro\m}$\times$\SI{50}{\micro\m} (fig. \ref{c5050_2e16}) since the geometry, and the drift distances for electrons and holes, will be the same).

 \section{Acknowledgements}
 
 This research has been partially financed by Spanish Ministry of Economy and Competitiveness through grant FPA2013-48308-C2-2-P and partially supported by the H2020 project AIDA-2020, GA no. 654168. 
 


  \bibliographystyle{elsarticle-num} 
    \bibliography{library_nima}

\end{document}